\documentclass[11pt,preprint,authoryear]{elsarticle}

\makeatletter
\def\ps@pprintTitle{%
 \let\@oddhead\@empty
 \let\@evenhead\@empty
 \def\@oddfoot{\centerline{\thepage}}%
 \let\@evenfoot\@oddfoot}
\makeatother

\usepackage{float}
\usepackage{a4wide}
\usepackage{amsmath}
\usepackage{amsfonts}
\usepackage{amssymb}
\usepackage{graphicx}
\usepackage{hyperref}
\usepackage{amsthm,kbordermatrix,blkarray}
\usepackage{cellspace}
\usepackage{color}

\usepackage{caption}
\usepackage{subfig}

\newtheorem{Def}{Definition}
\newtheorem{Prop}{Proposition}
\newproof{pf}{Proof}

\newtheorem{Rem}{\textit{Remark}}

\def\ee{{\mathbb E}}

\def\vv{{\mathbb V}}
\def\rr{{\mathbb R}}

\newcommand{\mb}[1]{\mathbf{#1}}

\makeatletter
\providecommand{\doi}[1]{%
  \begingroup
    \let\bibinfo\@secondoftwo
    \urlstyle{rm}%
    \href{http://dx.doi.org/#1}{%
      doi:\discretionary{}{}{}%
      \nolinkurl{#1}%
    }%
  \endgroup
}
\makeatother

\begin{document}

\begin{frontmatter}

\title{Adaptive use of replicated Latin Hypercube Designs for computing Sobol' sensitivity indices}

\author{Guillaume Damblin\corref{cor1}}
\ead{guillaume.damblin@cea.fr}
\cortext[cor1]{Corresponding author}

\author{Alberto Ghione}
\ead{alberto.ghione@cea.fr}

\address{CEA Saclay - DES/ISAS/DM2S/STMF - F-91191 Gif-sur-Yvette Cedex}

\begin{abstract}
As recently pointed out in the field of Global Sensitivity Analysis (GSA) of computer simulations, the use of replicated Latin Hypercube Designs (rLHDs) is a cost-saving alternative to regular Monte Carlo sampling to estimate first-order Sobol' indices. Indeed, two rLHDs are sufficient to compute the whole set of those indices regardless of the number of input variables. This relies on a permutation trick which, however, only works within the class of estimators called Oracle 2. In the present paper, we show that rLHDs are still beneficial to another class of estimators, called Oracle 1, which often outperforms Oracle 2 for estimating small and moderate indices. Even though unlike Oracle 2 the computation cost of Oracle 1 depends on the input dimension, the permutation trick can be applied to construct an averaged (triple) Oracle 1 estimator whose great accuracy is presented on a numerical example. 

Thus, we promote an adaptive rLHDs-based Sobol' sensitivity analysis where the first stage is to compute the whole set of first-order indices by Oracle 2. If needed, the accuracy of small and moderate indices can then be reevaluated by the averaged Oracle 1 estimators. This strategy, cost-saving and guaranteeing the accuracy of estimates, is applied to a computer model from the nuclear field.
\end{abstract}

\end{frontmatter}

\section{Introduction}

Engineering studies of physical systems based on computer simulations are often complemented by a stage of uncertainty quantification to assess the robustness of predictions of output Quantities of Interest (QoIs). A major source of uncertainty is the one coming from input variables whose values can be stochastic when the conditions of the system are not exactly known. In this context, Global Sensitivity Analysis (GSA) aims to identify the most influential variables on the output QoIs so that the effort to assess the uncertainty, and then to reduce it, can be put on those.

Over the years, a high number of scientific publications have investigated various GSA strategies. The most popular is that of Sobol' \citep{Sobol93}. The Sobol' method aims to compute indices measuring the variance contribution of each input variable and group of variables to the total variance of an output QoI. There are two sorts of indices that are commonly computed in priority: the first-order and total-order indices. The former gives the part of variance due to the individual effect of each input variable on the output while the latter measures a global effect of the variable including its possible interactions with the other inputs.


\medskip
Although computer speed and resources are growing up exponentially, computer models can still be computationally expensive for many applications. In this context, practitioners who aim to compute Sobol' indices have to opt for a strategy between the following two$:$ either constructing an emulator of the computer model, then calling it as much as necessary to compute estimates with a desired level of accuracy \citep{Ghanem16}, or computing indices directly from a small set of carefully chosen simulations. The first way can work well if the approximation error due to the replacement of the computer model by the emulator is low. The other way, which interests us in this work, asks a great effort in designing the experiments to maximize the amount of statistical information. To achieve this, replicated Latin Hypercube designs (rLHDs) presented for the first time in \citet{McKay99} are promising. 
Indeed, a permutation trick can be applied to the rows of such designs so that all the first-order Sobol' indices can be estimated with only two sets of simulations regardless of the input dimension \citep{Mara08}. Later on, the performances of rLHDs-based estimators have been studied on some academic examples and both their consistency and asymptotic normality have been proven \citep{Tissot15}. However, these estimators can only be computed through the class of Oracle 2 estimators. In the paper, we will address another class of estimators called Oracle 1 whose computation relies on one more set of simulations per index than Oracle 2 \citep{Owen13}. Despite this extra cost, the Oracle 1 estimators often have the advantage of being more accurate than Oracle 2 to estimate small and moderate indices \citep{Glen2012}. After clarifying this point with the help of the state of the art, the possibility of further improving the accuracy of rLHDs-based Oracle 1 estimators will be investigated. In this way, we will present an averaged (triple) Oracle 1 estimator whose accuracy outperforms that of the simple Oracle 1. 

The second contribution of the paper is to promote a two-stage strategy for computing Sobol' indices, which relies on rLHDs.  The motivation is to restrict as much as possible the number of simulations with respect to the complexity of the model (additive, non additive, etc). The first stage is to use Oracle 2 to estimate the whole set of first-order Sobol' indices using two sets of simulations only. The second stage is then to compute, if necessary, the averaged (triple) Oracle 1 estimators to refine the estimates returned by Oracle 2. As a bonus, the total-order Sobol' indices can be computed for the input variables whose first-order indices have been computed by Oracle 1.

\medskip
The paper is organized as follows$:$ Section $2$ recalls the mathematical definition of Sobol' indices. Section $3$ presents the two classes of estimators Oracle 1 and Oracle 2 which will be compared to each other in terms of the magnitude of their spurious correlation. Section $4$ consists of presenting rLHDs, the rLHDs-based Oracle 2 estimators and then introducing the new averaged (triple) rLHD-based Oracle 1 estimator. Section $5$ promotes an adaptive strategy for computing Sobol' indices which uses both the rLHDs-based Oracle 1 and 2 estimators. Section $6$ performs numerical comparisons of the estimators and tests the adaptive strategy in three possible situations. Section $7$ implements this strategy for sensitivity analysis of a computer model from the nuclear field. Section $8$ concludes the paper.

\section{Sobol' indices}

\medskip
Throughout the paper, the computer model is likened to a function $y(.)$ whose inputs are synthesized into the vector $\mb{x}=(x_1,\cdots,x_d) \in\mathcal{X}\subset\rr^{d}$. The corresponding output QoI $y(\mb{x})$ is assumed scalar$:$
\begin{equation}
\begin{alignedat}{2}
\label{model_notation}
y \colon & \mathcal{X} &\: &\to \rr \\
& \mb{x} & &\mapsto y(\mb{x}).
\end{alignedat}
\end{equation}
This function will be used as a "black-box" in the sense that the computation of Sobol' indices does not require the mathematical relation between $\mb{x}$ and $y(\mb{x})$. The input uncertainty is assessed by a random vector $X=(X_1,\cdots,X_d)$ whose marginal distributions $X_i$ are independent one another $(1\leq i\leq d)$. Hence, the probability measure of $X$ is equal to the product of marginals$:$
\begin{equation}
\label{inputs_measure}
\mu(\mb{x})=\mu(x_1)\,\otimes\mu(x_2)\,\cdots\otimes\, \mu(x_d).
\end{equation}
The output random variable $Y:=y(X)$ is obtained by propagating $X$ through the model and is assumed square-integrable ($\ee[Y^{2}]<\infty$). 

\subsection{The ANOVA decomposition}

Let $X_{\mb{u}}$ denote the random vector formed by the subset of variables $\{X_k\}_k$ with $k$ belonging to $\mb{u}\subseteq \{1,2,\cdots,d\}$. The starting point for defining Sobol' indices is the ANOVA (Analysis of variance) finite expansion of $Y$ \citep{Sobol93}, also known as HDMR (High-dimensional model representation)$:$
\begin{align}
\label{Hoeff}
Y=&\,\mu_Y+\sum_{i=1}^{d}y_i(X_i)+\sum_{1\leq i<j \leq d}^{d}y_{i,j}(X_i,X_j)+\cdots+\,\,y_{1,2,\cdots,d}(X_1,X_2,\cdots,X_d) \\
=&\,\mu_{Y}+\sum_{\mb{u}\subseteq\{1,\cdots,d\}}y_{\mb{u}}(X_{\mb{u}})
\end{align}
The following constraint ensures the expansion exists and is unique,
\begin{equation}
\label{condANOVA}
\int y_{\mb{u}}(x_{\mb{u}}) \otimes_{k\in\mb{u}} d\mu(x_k)=0\,\,\,\,\,\,\,\forall\,\mb{u}\subseteq\{1,\cdots,d\}. 
\end{equation}
It can be deduced that $\mu_Y=\ee[Y]$, 
\begin{equation}
y_i(X_i)=\ee[Y|X_i=x_i]-\mu_{Y},
\end{equation}
and, recursively,
\begin{equation}
\label{InteractionIndex}
y_{\mb{u}}(X_{\mb{u}})=\, \ee[Y|X_{\mb{u}}=\mb{x}_{\mb{u}}]-\sum_{\mb{v} \subsetneq  \mb{u}} y_{\mb{v}}(X_{\mb{v}}).
\end{equation}

Eq. (\ref{Hoeff}) leads to the decomposition of the variance of $Y$ as (group of) variable(s) contributions of increasing dimension, whose magnitude is assessed through Sobol' indices. 

\subsection{Closed and interaction-effect Sobol' indices}

\medskip
Based on the previous ANOVA decomposition, let us define $\forall\,\mb{u}\subseteq\{1,\cdots,d\}$$:$
\begin{equation}
D_{\mb{u}}=\sum_{\mb{v}\subseteq\mb{u}}\int y^{2}_{\mb{v}}(\mb{x}_{\mb{v}})\otimes_{k\in\mb{v}} d\mu(x_k)\,\,\,\,\,\,\,\textrm{and}\,\,\,\,\,\,\,D=\int_{\mathcal{X}}y^{2}(\mb{x})\textrm{d}\mu(\mb{x})-\mu_{Y}^2.
\end{equation}
\begin{Def}
The closed Sobol' index, denoted by $S_{\mb{u}}$, measures the proportion of the variance of $Y$ due to $X_{\mb{u}}$. It is written as
\begin{align}
\label{Sobol'_indices}
\nonumber
S_{\mb{u}}\,\,=&\,\,\frac{\vv[\ee(Y|X_{\mb{u}})]}{\vv[Y]} \\
=&\,\,\frac{D_{\mb{u}}}{D}\,\,\in[0,1]. \\
\end{align} 
\end{Def}

\noindent
When $\mb{u}=\{i\}$ we obtain the first-order Sobol' index$:$
\begin{equation}
\label{Sobol'1}
S_{\{i\}}=\frac{\vv[\ee(Y|X_i)]}{\vv[Y]}\,\,\,;\,\,\,1\leq i\leq d
\end{equation} 
which measures the proportion of the variance of $Y$ due to the uncertainty of $X_i$ alone. If $\mb{u}=\{i,j\}$ with $1\leq i\neq j\leq d$, we obtain the closed second-order Sobol' index of $(X_i,X_j)$ which is the sum of three indices$:$ the first-order index of $X_i$, the one of $X_j$ and an interaction index between $X_i$ and $X_j$ denoted by $\underline{S}_{\mb{u}}$. Thus,
\begin{equation}
\label{interact_ij}
\underline{S}_{\mb{u}}=S_{\mb{u}}-S_{\{i\}}-S_{\{j\}}=\frac{\vv[\ee(Y|X_i,X_j)]-\vv[\ee(Y|X_i)]-\vv[\ee(Y|X_j)]}{\vv[Y]}.
\end{equation} 
The index $\underline{S}_{\mb{u}}$ in Eq. (\ref{interact_ij}) measures the part of the variance of $Y$ due to the interaction of $X_i$ with $X_j$. The next definition extends such indices to measure interaction effects of higher orders.
\begin{Def}
The interaction Sobol' index $\underline{S}_{\mb{u}}$ of order $|\mb{u}|$, related to the subset of variables $X_{\mb{u}}$, measures the average proportion of $Y$'s variance due to the interaction effect between all the variables of $X_{\mb{u}}$. Thus,
\begin{align}
\underline{S}_{\mb{u}}=&\,S_{\mb{u}}-\sum_{\mb{v} \subsetneq  \mb{u}} \underline{S}_{\mb{v}}.
\end{align}
\end{Def}
The index $\underline{S}_{\mb{u}}$ is obtained by removing from $S_{\mb{u}}$ both all the interaction indices of lower order and the first-order indices. As a special case, $\underline{S}_{\mb{u}}=S_{\mb{u}}$ if $\mb{u}=\{i\}$. In addition, we have the important property that
\begin{equation}
\label{sum_var}
\sum_{\mb{u} \subseteq \{1,\cdots,d\}} \underline{S}_{\mb{u}}=1.
\end{equation}
The number of indices that makes up the sum is equal to $2^{d}-1$. When there are no interaction indices of any order, we have
\begin{equation}
\label{sum_var_part}
\sum_{i=1}^{d} S_{\{i\}}=1.
\end{equation}
The model is then additive as a sum of individual contribution of each variable $X_i$ to the total variance of $Y$.

\subsection{The total-order Sobol' indices}

\medskip
A total-order Sobol' index measures the proportion of $Y$'s variance due to a subset of variables $X_{\mb{u}}$ by accounting for not only the effect of $X_{\mb{u}}$ but also its possible interactions with the remaining variables $X_{-\mb{u}}:=X\setminus X_{\mb{u}}$.
\begin{Def}
The total-order Sobol' index of $X_{\mb{u}}$ is equal to
\begin{equation}
\label{ind_tot}
S_{\mb{u}}^{T}=S_{\mb{u}}+\sum_{\mb{v} \supsetneq \mb{u}} \underline{S}_{\mb{v}}.
\end{equation}
\end{Def}
A special case is the total-order Sobol' index of $X_i$, denoted by $S_{\{i\}}^{T}$. This index is thus equal to the sum of $S_{\{i\}}$ and the possible interaction Sobol' indices of every group of variables $X_{\mb{v}}\supsetneq X_i$$:$
\begin{align}
\label{sob_tot}
S_{\{i\}}^{T}=S_{\{i\}}+\sum_{\mb{v} \supsetneq \{i\}} \underline{S}_{\mb{v}}=&\,\,\frac{\vv[Y]-\vv[\ee(Y|X_{-\{i\}})]}{\vv[Y]}\\
=&\,\,\frac{\ee[\vv(Y|X_{-\{i\}})]}{\vv[Y]}.
\end{align}
with $X_{-\{i\}}$ being the vector of all variables except $X_{i}$.

\section{Estimators}

\label{section3}
In this section, we review some estimators of the Sobol' indices which can be seen as empirical correlation coefficients and whose accuracy will be assessed according to the magnitude of their spurious correlation. 

\medskip
\subsection{Estimation of first-order Sobol' indices}


\medskip
In the paper, we focus on two classes of estimators called Oracle $1$ and Oracle $2$, as named in \citet{Owen13}. Below $\mathcal{M}_{N,d}(\rr)$ refers to the real $N\times d$ matrices. Let $\mb{X}$ and $\mb{W}$ be two designs of experiments, each made up of $N$ samples generated from the probability distribution of $X$. Then, 
\begin{equation}
\label{Xdesign}
\mb{X}=[\mb{x}^{1},\cdots,\mb{x}^{N}]^{T}\,\,\,\,\,\,\in\,\,\,\,\,\,\mathcal{M}_{N,d}(\rr)
\end{equation}
where $\mb{x}^{k}=(x^{k}_1,\cdots,x^{k}_d)^{T}$ for $1\leq k\leq N$ and
\begin{equation}
\label{Wdesign}
\mb{W}=[\mb{w}^{1},\cdots,\mb{w}^{N}]^{T}\,\,\,\,\,\,\in\,\,\,\,\,\,\mathcal{M}_{N,d}(\rr)
\end{equation}
where $\mb{w}^{k}=(w^{k}_1,\cdots,w^{k}_d)^{T}$. The two sets of simulations run at $\mb{X}$ and $\mb{W}$ are respectively denoted by
\begin{equation}
x^k:=y(\mb{x}^k)\,\,\,\,\,\,\text{and}\,\,\,\,\,\,w^k:=y(\mb{w}^k).
\end{equation}
A third design, denoted by $\mb{W}_{-i}$, is then constructed as $\mb{W}$ except the $i$-th column is equal to that of $\mb{X}$$:$
\begin{equation}
\mb{W}_{-i}=[\mb{w}_{-i}^1,\cdots, \mb{w}^{N}_{-i}]^{T}\,\,\,\,\,\,\in\,\,\,\,\,\,\mathcal{M}_{N,d}(\rr).
\end{equation} 
where $\mb{w}_{-i}^{k}=(w^{k}_1,\cdots,w^{k}_{i-1},x^{k}_{i},w^{k}_{i+1},\cdots,w^{k}_d)^{T}$.
The simulations run at $\mb{W}_{-i}$ are then denoted by
%

\medskip
\begin{equation} 
w_{-i}^k:=y(\mb{w}_{-i}^{k}).
\end{equation}

\subsubsection{The Oracle 2 class of estimators}

\vspace{0.4cm}
Eq. (\ref{Sobol'1}) can be rewritten as the regression coefficient below$:$
\begin{equation}
S_{\{i\}}=\,\,\frac{\text{Cov}\big(y(X),y(W_{-i})\big)}{\sigma_Y^2}
\end{equation}
with $\sigma_Y^2:=\vv[Y]$. The distribution of $y(W_{-i})$ is the same as that of $Y=y(X)$, so $S_{\{i\}}$ comes down to the Pearson correlation coefficient between $y(X)$ and $y(W_{-i})$ \citep{Martinez11}$:$
\begin{equation}
S_{\{i\}}=\,\,\rho\big(y(X),y(W_{-i})\big).
\label{SiCorr}
\end{equation}
If an oracle could provide the true mean $\mu_{Y}$ and variance $\sigma_Y^2$, the estimator of $S_{\{i\}}$ proposed in \citet{Owen13} would be computed as
\begin{equation}
\label{SiCorr_estim_oracle}
\hat{S}_{\{i\}}=\frac{\sum_{k=1}^{N}(x^k-\mu_{Y})(w_{-i}^k-\mu_{Y})}{N\sigma_Y\times\sigma_Y}.
\end{equation}
In real studies, $\mu_{Y}$ and $\sigma_Y^2$ are unknown and should be estimated. A first achievable approximation of (\ref{SiCorr_estim_oracle}) is the empirical Pearson correlation coefficient \citep{Glen2012}$:$
\begin{equation}
\label{SiCorr_estim}
\hat{S}_{\{i\}}=\frac{\sum_{k=1}^{N}(x^k-\bar{x})(w_{-i}^k-\bar{w}_{-i})}{\sqrt{\sum_{k=1}^{N}(x^k-\bar{x})^2} \sqrt{\sum_{k=1}^{N}(w_{-i}^k-\bar{w}_{-i})^2}} 
\end{equation}
with $\bar{x}$ and $\bar{w}_{-i}$ being the empirical mean of the simulations run at $\mb{X}$ and $\mb{W}_{-i}$ respectively. Another way is to pool all these simulations to compute improved estimates of $\ee[Y]$ and $\vv[Y]$, leading to
\begin{equation}
\label{Oracle2_janon}
\hat{S}_{\{i\}}=\,\,\frac{\sum_{k=1}^{N}(x^k-\hat{\mu}_{Y})(w_{-i}^k-\hat{\mu}_{Y})}{N\hat{\sigma}_Y^2}
\end{equation}
in which
\begin{equation}
\label{improved_mu_sigma2}
\hat{\sigma}_Y^2=\frac{1}{N}\sum_{k=1}^{N} \Big(\frac{(x^{k})^{2}+(w_{-i}^{k})^{2}}{2}-\hat{\mu}^2_Y\Big)
\,\,\,\,\,\text{and} \,\,\,\,\,\hat{\mu}_Y=\frac{\bar{x}+\bar{w}_{-i}}{2}.
\end{equation}
It was proven in \citet{Janon13} that Eq. (\ref{Oracle2_janon}) gives a consistent and asymptotically Gaussian estimator of $S_{\{i\}}$. The next equation rewrites this estimator as the empirical correlation coefficient between the simulations run at $\mb{X}$ and those run at $\mb{W}_{-i}$ after having standardized them with respect to $\hat{\mu}_Y$ and $\hat{\sigma}_Y$$:$
\begin{equation}
\label{Oracle2_janon_corr}
\hat{S}_{\{i\}}=\,\,\frac{1}{N}\sum_{k=1}^{N}\Big(\frac{x^k-\hat{\mu}_{Y}}{\hat{\sigma}_Y} \Big)\times \Big(\frac{w_{-i}^k-\hat{\mu}_{Y}}{\hat{\sigma}_Y} \Big).
\end{equation}
Once $N$ is large enough, Eqs. (\ref{SiCorr_estim}) and (\ref{Oracle2_janon_corr}) yield close estimates.

\subsubsection{The Oracle 1 class of estimators}

\vspace{0.4cm}
As introduced in \citet{Owen13}, the Oracle 1 estimator is computed as
\begin{equation}
\label{SiCorr_estim_oracle1}
\hat{S}_{\{i\}}=\frac{\sum_{k=1}^{N}(x^k-\mu_{Y})(w_{-i}^k-w^{k})}{N\sigma_Y^2},
\end{equation}
which requires running three sets of simulations at $\mb{X}$, $\mb{W}$ and $\mb{W}_{-i}$. Interestingly, this estimator matches the one below proposed in \citet{Glen2012} :
\begin{equation}
\label{SiCorr_estim_oracle1_expand}
\hat{S}_{\{i\}}=\frac{\sum_{k=1}^{N} (x^k-\mu_Y)(w_{-i}^k-\mu_Y) - \sum_{k=1}^{N} (x^k-\mu_Y)(w^{k}-\mu_Y)}{N\sigma^2_{Y}}
\end{equation}
Indeed, factoring by $x^k-\mu_Y$ matches Eq. (\ref{SiCorr_estim_oracle1}).  Being the empirical correlation coefficient between two independent set of simulations, the second summation in Eq. (\ref{SiCorr_estim_oracle1_expand}) tends to $0$ as $N$ increases, which proves that Oracle 1 converges to $S_{\{i\}}$. 

As $y(W)$, $y(X)$ and $y(W_{-i})$ have the same distribution, the three sets of simulations run at $\mb{X}$, $\mb{W}$ and $\mb{W}_{-i}$ can be pooled to estimate both $\mu_{Y}$ and $\sigma_Y^{2}$. An efficient practical approximation of Eq. (\ref{SiCorr_estim_oracle1}) is then given by
\begin{equation}
\label{oracle1}
\hat{S}_{\{i\}}=\frac{\sum_{k=1}^{N} (x^k-\hat{\mu}_Y)(w_{-i}^k-w^k)}{N\hat{\sigma}_Y^2}
\end{equation}
with
\begin{equation}
\hat{\sigma}_Y^2=\frac{1}{N}\sum_{k=1}^{N} \Big(\frac{(x^{k})^{2}+(w_{-i}^{k})^{2}+(w^{k})^{2}}{3}-\hat{\mu}_Y\Big)
\,\,\,\,\text{and} \,\,\,\,\hat{\mu}_Y=\frac{\bar{x}+\bar{w}+\bar{w}_{-i}}{3}.
\end{equation}
Table \ref{tablecost} recaps the number of simulations needed respectively by Oracle 2 and Oracle 1 to compute $S_{\{i\}}$. 

\begin{table}[H]
\setlength\extrarowheight{2pt}
\vspace{0.3cm}
\centering
\begin{tabular}{|l|l|l|}
\hline
Estimator's name & cost by index & total cost \\
\hline
Oracle 1 & $3N$ & $N(d+2)$ \\
\hline
Oracle 2 & $2N$ & $N(d+1)$ \\
\hline
\end{tabular}
\caption{The middle and last columns present respectively the number of simulations for estimating one first-order index and all first-order indices.}
\label{tablecost}
\end{table}

\begin{Rem}
Although Sobol' indices take positive values as ratios of variances, negative estimates can sometimes be yielded when $S_{\{i\}}$ is close to $0$. This is because the Oracle 1 and Oracle 2 estimators are written as correlation coefficients which do not ensure positive estimates due to statistical variability.
\end{Rem}

\subsection{An estimator of the total-order Sobol' index $S_{\{i\}}^{T}$}

\medskip
For $1\leq i\leq d$, the total-order index of $X_i$ is equal to
\begin{align}
S_{\{i\}}^{T}=&\,1-S_{\{-i\}}\\
=&\,1-\frac{\vv[\ee(Y|X_{\{-i\}})]}{\vv[Y]}\\
=&\,1-\frac{\text{Cov}(Y(W),Y(W_{-i}))}{\sigma_Y^2}.
\end{align}
An estimator of $S_{\{i\}}^{T}$ can then be computed as
\begin{equation}
\label{TotalIndexEstim}
S_{\{i\}}^{T}=1-\frac{\sum_{k=1}^{N} (w^k-\hat{\mu}_Y)(w_{-i}^k-\hat{\mu}_Y)}{N\hat{\sigma}^2_{Y}}
\end{equation}
with
\begin{equation}
\hat{\sigma}_Y^2=\frac{1}{N}\sum_{k=1}^{N} \Big(\frac{(w_{-i}^{k})^{2}+(w^{k})^{2}}{2}-\hat{\mu}_Y\Big)
\,\,\,\,\text{and} \,\,\,\,\hat{\mu}_Y=\frac{\bar{w}+\bar{w}_{-i}}{2}.
\end{equation}
This estimator of $S_{\{i\}}^{T}$ needs no extra simulation as compared to Oracle 1.

\subsection{Comparison of estimators}

\label{sec_comp}

\medskip
A comparative study of the variance related to the different estimators of closed Sobol' indices has been presented in \citet{Owen13}. Several rules concerning Oracle 1 and Oracle 2 have been established$:$
\begin{enumerate}[1.]
\item Oracle 2 is better to estimate large $S_{\{i\}}$,

\smallskip
\item Oracle 2 is better to estimate $S_{\{i\}}$ when $\sigma_Y^2$ is mostly explained by the interaction index between $X_i$ and the other variables,

\smallskip
\item  Oracle 1 is better when both $S_{\{i\}}$ and $S^{T}_{\{i\}}$ are small,

\smallskip
\item  No definitive rule can be provided when $S_{\{i\}}$ is small or moderate but not $S^{T}_{\{i\}}$, 
\end{enumerate}
For the item 4), a rule was given by \citet{Glen2012} that Oracle $1$ would be more accurate than Oracle 2 if $S_{\{i\}}^{T}<0.5$. The arguments behind this rule are provided hereafter.

\medskip
Let us denote$:$
\begin{itemize}
\item $\mb{X}(i)$ and $\mb{W}_{-i}(i)$ respectively the $i$-th column of $\mb{X}$ and $\mb{W}_{-i}$,
\item $\mb{X}(-i)$ and $\mb{W}_{-i}(-i)$ respectively all the columns of $\mb{X}$ and $\mb{W}_{-i}$ except the $i$-th. 
\end{itemize}
According to Eq. (\ref{SiCorr}), the Sobol' index $S_{\{i\}}$ measures the correlation between the output variables $Y=y(\mb{X})$ and $y(\mb{W}_{-i})$ induced by the agreement between the columns $\mb{X}(i)$ and $\mb{W}_{-i}(i)$. However, part of this correlation can also result from the non-zero empirical correlation between the remaining columns $\mb{X}(-i)$ and $\mb{W}_{-i}(-i)$. Although converging to $0$ with variance about equal to $N^{-1}$, this spurious correlation can significantly taint the accuracy of estimators when $N$ is small or even moderate. According to \citet{Glen2012}, the spurious correlation of Oracle 2 is proportional to
\begin{equation}
\label{spurious2}
S_{\{-i\}}+\underline{S}_{\{i,-i\}}
\end{equation} 
while the spurious correlation of Oracle 1 is proportional to
\begin{equation}
\label{spurious1}
S_{\{i\}}+2\underline{S}_{\{i,-i\}}.
\end{equation} 
By comparing (\ref{spurious1}) to (\ref{spurious2}), Glen and Isaacs deduced that the spurious correlation of Oracle 1 is lower than that of Oracle 2 if and only if
\begin{equation}
\label{condition}
S_{\{i\}}^{T}<\frac{1}{2}.
\end{equation}
We show in \ref{appendix} that the above rule is actually not valid. However, in absence of any interaction between $X_i$ and the other variables, this condition actually holds and becomes
\begin{equation}
\label{conditionGlen}
S_{\{i\}}<\frac{1}{2}.
\end{equation}

\subsection{Complexity of models} 

\label{complexity}

Recall that numerical models can be qualitatively ranked according to their degree of complexity \citep{Kuch2009MC}$:$
\begin{itemize}
\item Type A models where a few variables are dominant through the value of their first-order index,
\item Type B models containing low-order interactions between variables,
\item Type C models containing significant high-order interactions between variables.
\end{itemize}

Type A models are additive or close whereas Type B and C models are made up of interaction effects, more or less numerous.

\section{Replicated Latin Hypercube Designs (rLHDs)}

\medskip
The estimators presented in Section \ref{section3} are based on the input designs $\mb{X}$, $\mb{W}$ and $\mb{W}_{-i}$. The way of generating them can have a strong impact on the accuracy of estimates. In this section, we present replicated Latin Hypercubes which are highly cost-saving designs to estimate the whole set of first-order Sobol' indices by Oracle 2.

\subsection{Definitions}

\medskip
\begin{Def}
Let $\Pi^{N}$ be the permutation group of the set of integers $\{1,2,\cdots,N\}$. A matrix $\mb{X}\in\mathcal{M}_{N,d}(\rr)$ is a Latin Hypercube Design (LHD) if each of its columns is constructed as a permutation $\pi_i\in\Pi^{N}$.
\end{Def}

\begin{Def}
\label{replicatedDef}
Two designs $\mb{X}, \mb{W}\in\mathcal{M}_{N,d}(\rr)$ are replicated to each other if there is a permutation $\pi_i$ such that $\pi_i\big(\mb{W}(i)\big)=\mb{X}(i)$ with $\mb{X}(i)$ and $\mb{W}(i)$ being the $i$-th column of $\mb{X}$ and $\mb{W}$ respectively.
\end{Def}

The structure of LHDs makes them automatically replicated, as shown below.
\paragraph{Example}
Let $\mb{X}, \mb{W}\in\mathcal{M}_{8,2}(\rr)$ be two independent LHDs such that
\begin{equation}
\mb{X}=
\begin{blockarray}{rcc}
      & \mb{X}(1) & \mb{X}(2) \\
  \begin{block}{r[cc]}
  & 1 & 4 \\
  & 3 & 5 \\
  & 8 & 6 \\
  & 4 & 7 \\
  & 6 & 1 \\
  & 2 & 8 \\
  & 5 & 3 \\
  & 7 & 2 \\
  \end{block}
\end{blockarray}
\quad
\mb{W}=
\begin{blockarray}{rcc}
      & \mb{W}(1) & \mb{W}(2) \\
  \begin{block}{r[cc]}
  & 3 & 1 \\
  & 1 & 2 \\
  & 4 & 8 \\
  & 5 & 4 \\
  & 8 & 7 \\
  & 2 & 6 \\
  & 7 & 3 \\
  & 6 & 5 \\
  \end{block}
\end{blockarray}
\label{twolhd}
\end{equation}
The columns $\mb{X}(1)$ and $\mb{X}(2)$ match the permutations $\pi^{X}_1=(238756)$ and $\pi^{X}_2=(14736825)$ respectively while the columns $\mb{W}(1)$ and $\mb{W}(2)$ match the permutations $\pi^{W}_1=(1345862)$ and $\pi^{W}_2=(7385)$ respectively. If the permutation $ \pi_1:=\pi^{X}_1\circ(\pi^{W}_1)^{-1}$ is applied to the rows of $\mb{W}$, then 
\begin{equation}
\pi_1\big(\mb{W}(1)\big)=\mb{X}(1).
\end{equation}
Similarly, if the permutation $\pi_2:=\pi^{X}_2\circ(\pi^{W}_2)^{-1}$ is applied to the rows of $\mb{W}$, then 
\begin{equation}
\pi_2\big(\mb{W}(2)\big)=\mb{X}(2). 
\end{equation}
According to Definition \ref{replicatedDef}, $\mb{X}$ and $\mb{W}$ are thus replicated to each other.

\vspace{0.5cm}
\begin{Def}
\label{randomizedLHD}
Assume that every marginal $X_i$ follows a uniform distribution on $[0,1]$ for $1\leq i\leq d$. 
Let $\{\pi^X_1,\pi^X_2,\cdots,\pi^X_d\}$ be a set of $d$ independent permutations of $\Pi^{N}$. For $1\leq k\leq N$, generate the uniform samples$:$
\begin{equation}
\label{PerturbUnif}
U_{i,k}\underset{i.i.d.}\thicksim\mathcal{U}\left(-\frac{1}{2},\frac{1}{2}\right).
\end{equation}
Then, the design
\begin{equation}
\mb{X}=\Big[\frac{\pi^X_1-0.5+U_{1,\pi^X_1}}{N},\cdots,\frac{\pi^X_i-0.5+U_{i,\pi^X_i}}{N},\cdots,\frac{\pi^X_d-0.5+U_{d,\pi^X_d}}{N} \Big]\in \mathcal{M}_{N,d}([0,1])
\label{XWrlhd}
\end{equation}
is called a randomized LHD\footnote{A randomized LHD is sometimes called Latin Hypercube Sampling (LHS)}, such that for $1\leq i\leq d$$:$
\begin{equation}
\frac{\pi^X_i-0.5+U_{i,\pi^X_i}}{N}=\Big(\frac{\pi^X_i(1)-0.5+U_{i,\pi^X_i(1)}}{N},\cdots,\frac{\pi^X_i(N)-0.5+U_{i,\pi^X_i(N)}}{N}\Big)^{T}\in\rr^{N}.
\end{equation}
\end{Def}

\begin{Prop}
Let $\mb{X}$ and $\mb{W}$ be two randomized LHDs such that$:$
\begin{itemize}
\item they are constructed with the two sets of mutually independent permutations $\{\pi^X_1,\pi^X_2,\cdots,\pi^X_d\}$ and $\{\pi^{W}_1,\pi^{W}_2,\cdots,\pi^{W}_d\}$,
\item they share the same realizations $U_{i,k}$ for $1\leq i\leq d$ and $1\leq k\leq N$.
\end{itemize}
Then, $\mb{X}$ and $\mb{W}$ are replicated to each other.
\end{Prop}

\begin{figure}[h]
\centering
\includegraphics[scale=0.5]
{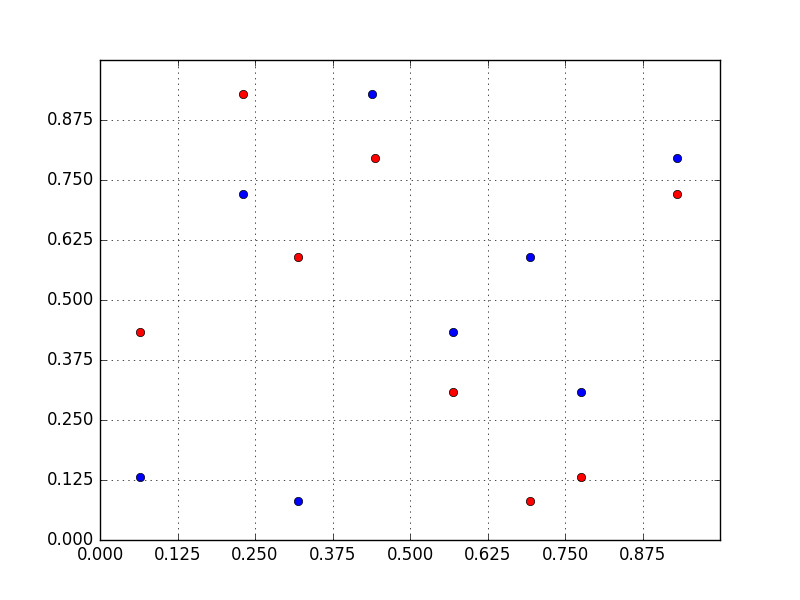}
\caption{Two randomized replicated LHDs $\mb{X}$ (red) and $\mb{W}$ (blue)}
\label{XWrlhd_randomized_fig}
\end{figure}

Eq. (\ref{XWrlhd}) transforms rLHDs into randomized rLHDs. Applying it to both designs of Eq. (\ref{twolhd}) gives the two randomized rLHDs $\mb{X}$ and $\mb{W}$ presented in Eq. (\ref{XWrlhd_randomized}) below. By denoting $U^{\prime}_{i,k}:=U_{i,k}/8$, then

\begin{equation}
\label{XWrlhd_randomized}
\mb{X}=
\begin{blockarray}{rcc}
      & \mb{X}(1) & \mb{X}(2) \\
  \begin{block}{r[cc]}
 & 0.0625+U^{\prime}_{1,1} & 0.4375+U^{\prime}_{2,4} \\
 & 0.3125+U^{\prime}_{1,3} & 0.5625+U^{\prime}_{2,5} \\
 & 0.9375+U^{\prime}_{1,8} & 0.6875+U^{\prime}_{2,6} \\
 & 0.4375+U^{\prime}_{1,4} & 0.8125+U^{\prime}_{2,7} \\
 & 0.6875+U^{\prime}_{1,6} & 0.0625+U^{\prime}_{2,1} \\
 & 0.1875+U^{\prime}_{1,2} & 0.9375+U^{\prime}_{2,8} \\
 & 0.5625+U^{\prime}_{1,5} & 0.3125+U^{\prime}_{2,3} \\
 & 0.8125+U^{\prime}_{1,7} & 0.1875+U^{\prime}_{2,2} \\
  \end{block}
\end{blockarray}
\quad
\mb{W}=
\begin{blockarray}{rcc}
      & \mb{W}(1) & \mb{W}(2) \\
  \begin{block}{r[cc]}
 & 0.3125+U^{\prime}_{1,3} & 0.0625+U^{\prime}_{2,1} \\
 & 0.0625+U^{\prime}_{1,1} & 0.1875+U^{\prime}_{2,2} \\
 & 0.4375+U^{\prime}_{1,4} & 0.9375+U^{\prime}_{2,8} \\
 & 0.5625+U^{\prime}_{1,5} & 0.4375+U^{\prime}_{2,4} \\
 & 0.9375+U^{\prime}_{1,8} & 0.8125+U^{\prime}_{2,7} \\
 & 0.1875+U^{\prime}_{1,2} & 0.6875+U^{\prime}_{2,6} \\
 & 0.8125+U^{\prime}_{1,7} & 0.3125+U^{\prime}_{2,3} \\
 & 0.6875+U^{\prime}_{1,6} & 0.5625+U^{\prime}_{2,5} \\
  \end{block}
\end{blockarray}
\end{equation}
Fig. \ref{XWrlhd_randomized_fig} displays them for a specific set of realizations $U_{i,k}$ ($1\leq i\leq 2$ and $1\leq k \leq 8$).

\begin{Rem}
\label{inverseCDF}
If the probability distribution of $X_i$ is not uniform on $[0,1]$, the inverse transformation method needs to be applied to the $i$-th column of the randomized rLHDs in Definition \ref{randomizedLHD}.
\end{Rem}

\subsection{The rLHDs for the computation of first-order Sobol' indices}

\label{perm_trick}

\medskip
McKay proposed to estimate the whole set of first-order Sobol' indices with $rN$ replicated LHDs \citep{McKay99}. The total number of simulations is independent of $d$, but the number of replications $r$ to achieve a good level of accuracy is unclear. Moreover, the authors used a different estimator as Oracle 1 and 2.

Mara and Joseph put in evidence that two replicated designs (not necessarily LHDs) are enough to compute the whole set of first-order Sobol' indices by Oracle 2 \citep{Mara08}. This is actually possible because the matrix obtained by reordering the rows of $\mb{W}$ with the permutation $\pi_i:=\pi^{X}_i\circ(\pi^{W}_i)^{-1}$ looks like $\mb{W}_{-i}$ ($1\leq i\leq d$). Below are the designs $\mb{W}_{-1}$ and $\mb{W}_{-2}$ derived from $\mb{W}$ in Eq. (\ref{XWrlhd_randomized}).

\begingroup\makeatletter\def\f@size{10}\check@mathfonts
\begin{equation}
\mb{W}_{-1}=
\begin{blockarray}{rcc}
      & \mb{W}_{-1}(1) & \mb{W}_{-1}(2) \\
  \begin{block}{r[cc]}
 & 0.0625+U^{\prime}_{1,1} & 0.1875+U^{\prime}_{2,2} \\
 & 0.3125+U^{\prime}_{1,3} & 0.0625+U^{\prime}_{2,1} \\
 & 0.9375+U^{\prime}_{1,8} & 0.8125+U^{\prime}_{2,7} \\
 & 0.4375+U^{\prime}_{1,4} & 0.9375+U^{\prime}_{2,8} \\
 & 0.6875+U^{\prime}_{1,6} & 0.5625+U^{\prime}_{2,5} \\
 & 0.1875+U^{\prime}_{1,2} & 0.6875+U^{\prime}_{2,6} \\
 & 0.5625+U^{\prime}_{1,5} & 0.4375+U^{\prime}_{2,4} \\
 & 0.8125+U^{\prime}_{1,7} & 0.3125+U^{\prime}_{2,3} \\
  \end{block}
\end{blockarray}
\quad
\mb{W}_{-2}=
\begin{blockarray}{rcc}
      & \mb{W}_{-2}(1) & \mb{W}_{-2}(2) \\
  \begin{block}{r[cc]}
 & 0.5625+U^{\prime}_{1,5} & 0.4375+U^{\prime}_{2,4} \\
 & 0.6875+U^{\prime}_{1,6} & 0.5625+U^{\prime}_{2,5} \\
 & 0.1875+U^{\prime}_{1,2} & 0.6875+U^{\prime}_{2,6} \\
 & 0.9375+U^{\prime}_{1,8} & 0.8125+U^{\prime}_{2,7} \\
 & 0.3125+U^{\prime}_{1,3} & 0.0625+U^{\prime}_{2,1} \\
 & 0.4375+U^{\prime}_{1,4} & 0.9375+U^{\prime}_{2,8} \\
 & 0.8125+U^{\prime}_{1,7} & 0.4375+U^{\prime}_{2,4} \\
 & 0.0625+U^{\prime}_{1,1} & 0.1875+U^{\prime}_{2,2} \\
  \end{block}
\end{blockarray}
\end{equation}
\endgroup

\medskip
The larger $d$, the most cost-saving the method. A theoretical study has proven that the Oracle 2 estimator (\ref{Oracle2_janon}) computed with two randomized rLHDs is still both consistent and asymptotically Gaussian \citep{Tissot15}. 

\subsection{On the use of rLHDs for \textit{Oracle 1}}

\label{secOracle1rLHDs}

\medskip
The reordering trick can still be leveraged to compute an improved rLHD-based Oracle 1 estimator. Let $\mb{W}(-i)$ and $\mb{W}_{-i}(-i)$ denote, respectively, the matrix $\mb{W}$ and $\mb{W}_{-i}$ apart from their $i$-th column. From Section \ref{section3}, $\mb{W}(-i)$ matches $\mb{W}_{-i}(-i)$ by construction. This is not true when using rLHDs because $\mb{W}_{-i}(-i)$ is obtained by the reordering trick. Another design, denoted by $\mb{Z}_i$, is thus needed for computing Oracle 1. Let $\mb{Z}_i(i)$ and $\mb{Z}_i(-i)$ denote respectively the $i$-th column of $\mb{Z}_i$ and all the others such that
\begin{itemize}
\item $\mb{Z}_i(-i)$ matches $\mb{W}_{-i}(-i)$,
\item $\mb{Z}_i(i)$ is created with a permutation $\pi_i^{Z_i}\in\Pi^{N}$ which is independent of both $\pi_i^{X}$ and $\pi_i^{W}$, but uses the same uniform samples $\{U_{i,k}\}$ as those of $\mb{X}$ and $\mb{W}$. 
\end{itemize}
Then, $\mb{X}$, $\mb{W}$ and $\mb{Z}_i$ are rLHDs one another. Based on the simulations run at these designs, $S_{\{i\}}$ can be estimated by
\begin{equation}
\label{oracle1_rhld1}
\hat{S}_{\{i\}}=
\frac{1}{\hat{\sigma}_Y^2N}\sum_{k=1}^{N} (x^k-\hat{\mu}_Y) \times
\big(w_{-i}^k-z_i^{k}\big) \\
\end{equation} 
with respect to the designs
\begin{align}
\overbrace{
\left[
\begin{array}{c|c}
\mb{X}(-i) & \mb{X}(i) \\
\end{array}
\right]}^{\mb{X}}
\Bigg|
\overbrace{
\left[
\begin{array}{c|c}
\mb{W}_{-i}(-i) & \mb{X}(i) \\
\end{array}
\right]}^{\mb{W}_{-i}}
\,\,
\overbrace{
\left[
\begin{array}{c|c}
\mb{W}_{-i}(-i) & \mb{Z}_i(i) \\
\end{array}
\right]}^{\mb{Z}_i}.
\label{firstOracle1}
\end{align}
By applying the permutation $\pi_i:=\pi^{Z_i}_i\circ (\pi_i^{X})^{-1}$ to the rows of $\mb{X}$, a new design denoted by $\mb{\tilde{X}}$ is generated whose $i$-th column matches that of $\mb{Z}_i$. Another Oracle 1 estimator can thus be computed as
\begin{equation}
\label{oracle1_rhld2}
\hat{S}_{\{i\}}=
\frac{1}{\hat{\sigma}_Y^2 N}\sum_{k=1}^{N} (\tilde{x}^k-\hat{\mu}_Y) \times
\big(z_i^{k}-w_{-i}^k\big) \\
\end{equation}
with respect to the designs
\begin{align}
\overbrace{
\left[
\begin{array}{c|c}
\mb{\tilde{X}}(-i) & \mb{Z}(i) \\
\end{array}
\right]}^{\mb{\tilde{X}}}
\Bigg|
\overbrace{
\left[
\begin{array}{c|c}
\mb{W}_{-i}(-i) & \mb{Z}_i(i) \\
\end{array}
\right]}^{\mb{Z}_i}
\,\,
\overbrace{
\left[
\begin{array}{c|c}
\mb{W}_{-i}(-i) & \mb{X}(i) \\
\end{array}
\right]}^{\mb{W}_{-i}}.
\label{secondOracle1}
\end{align}
The simulations run at $\mb{\tilde{X}}$ match those run at $\mb{X}$ through the reordering trick. By applying the permutation $\pi_i:=\pi^{Z_i}_i\circ (\pi_i^{W_{-i}})^{-1}$ to the rows of $\mb{W}_{-i}$, a new design denoted by $\widetilde{\mb{W}}_{-i}$ is generated whose $i$-th column matches that of $\mb{Z}_{i}$. Hence, a third estimator of Oracle 1 can be computed as
\begin{equation}
\label{oracle1_rhld3}
\hat{S}_{\{i\}}=
\frac{1}{\hat{\sigma}_Y^2N}\sum_{k=1}^{N} (\tilde{w}_{-i}^k-\hat{\mu}_Y) \times
\big(z_i^{k}-w_{-i}^k\big) \\
\end{equation}
with respect to the designs 
\begin{align}
\overbrace{
\left[
\begin{array}{c|c}
\mb{\widetilde{W}}(-i) & \mb{Z}(i) \\
\end{array}
\right]}^{\widetilde{\mb{W}}_{-i}}
\Bigg|
\overbrace{
\left[
\begin{array}{c|c}
\mb{W}_{-i}(-i) & \mb{Z}_i(i) \\
\end{array}
\right]}^{\mb{Z}_i}
\,\
\overbrace{
\left[
\begin{array}{c|c}
\mb{W}_{-i}(-i) & \mb{X}(i) \\
\end{array}
\right]}^{\mb{W}_{-i}}.
\label{thirdOracle1}
\end{align}
The simulations run at $\mb{\widetilde{W}}$ match those run at $\mb{W}_{-i}$. Finally, the three estimators can be put together to provide an averaged (triple) Oracle 1 estimator$:$
\begin{equation}
\label{oracle1triple}
\hat{S}_{\{i\}}=\frac{1}{3}\big[(\ref{oracle1_rhld1}) + (\ref{oracle1_rhld2}) + (\ref{oracle1_rhld3})\big].
\end{equation}

\vspace{0.4cm}
\paragraph*{\textbf{Example}} Let us compute an averaged (triple) Oracle 1 estimator of $S_{\{1\}}$. This requires a design $\mb{Z}_1$. Let $\pi_1^{Z_1}=(42371865)\in\Pi^{8}$ be a permutation generated randomly. The designs presented in Eq. (\ref{firstOracle1}) are then equal to
\begingroup\makeatletter\def\f@size{8.5}\check@mathfonts
\begin{align}
\overbrace{
\left[
\begin{array}{c|c}
0.0625+U^{\prime}_{1,1} & 0.4375+U^{\prime}_{2,4} \\
0.3125+U^{\prime}_{1,3} & 0.5625+U^{\prime}_{2,5} \\
0.9375+U^{\prime}_{1,8} & 0.6875+U^{\prime}_{2,6} \\
0.4375+U^{\prime}_{1,4} & 0.8125+U^{\prime}_{2,7} \\
0.6875+U^{\prime}_{1,6} & 0.0625+U^{\prime}_{2,1} \\
0.1875+U^{\prime}_{1,2} & 0.9375+U^{\prime}_{2,8} \\
0.5625+U^{\prime}_{1,5} & 0.3125+U^{\prime}_{2,3} \\
0.8125+U^{\prime}_{1,7} & 0.1875+U^{\prime}_{2,2} \\
\end{array}
\right]}^{\mb{X}}
\overbrace{
\left[
\begin{array}{c|c}
0.0625+U^{\prime}_{1,1} & 0.1875+U^{\prime}_{2,2} \\
0.3125+U^{\prime}_{1,3} & 0.0625+U^{\prime}_{2,1} \\
0.9375+U^{\prime}_{1,8} & 0.8125+U^{\prime}_{2,7} \\
0.4375+U^{\prime}_{1,4} & 0.9375+U^{\prime}_{2,8} \\
0.6875+U^{\prime}_{1,6} & 0.5625+U^{\prime}_{2,5} \\
0.1875+U^{\prime}_{1,2} & 0.6875+U^{\prime}_{2,6} \\
0.5625+U^{\prime}_{1,5} & 0.4375+U^{\prime}_{2,4} \\
0.8125+U^{\prime}_{1,7} & 0.3125+U^{\prime}_{2,3} \\
\end{array}
\right]}^{\mb{W}_{-1}}
\overbrace{
\left[
\begin{array}{c|c}
0.4375+U^{\prime}_{1,4} & 0.1875+U^{\prime}_{2,2} \\
0.1875+U^{\prime}_{1,2} & 0.0625+U^{\prime}_{2,1} \\
0.3125+U^{\prime}_{1,3} & 0.8125+U^{\prime}_{2,7} \\
0.8125+U^{\prime}_{1,7} & 0.9375+U^{\prime}_{2,8} \\
0.0625+U^{\prime}_{1,1} & 0.5625+U^{\prime}_{2,5} \\
0.9375+U^{\prime}_{1,8} & 0.6875+U^{\prime}_{2,6} \\
0.6875+U^{\prime}_{1,6} & 0.4375+U^{\prime}_{2,4} \\
0.5625+U^{\prime}_{1,5} & 0.3125+U^{\prime}_{2,3} \\
\end{array}
\right]}^{\mb{Z}_1}
\end{align}
\endgroup
The designs presented in Eq. (\ref{secondOracle1}) are then equal to
\begingroup\makeatletter\def\f@size{8.5}\check@mathfonts
\begin{align}
\overbrace{
\left[
\begin{array}{c|c}
0.4375+U^{\prime}_{1,4} & 0.8125+U^{\prime}_{2,7} \\
0.1875+U^{\prime}_{1,2} & 0.9375+U^{\prime}_{2,8} \\
0.3125+U^{\prime}_{1,3} & 0.5625+U^{\prime}_{2,5} \\
0.8125+U^{\prime}_{1,7} & 0.1875+U^{\prime}_{2,2} \\
0.0625+U^{\prime}_{1,1} & 0.4375+U^{\prime}_{2,4} \\
0.9375+U^{\prime}_{1,8} & 0.6875+U^{\prime}_{2,6} \\
0.6875+U^{\prime}_{1,5} & 0.0625+U^{\prime}_{2,1} \\
0.5625+U^{\prime}_{1,5} & 0.3125+U^{\prime}_{2,3} \\
\end{array}
\right]}^{\mb{\tilde{X}}}
\overbrace{
\left[
\begin{array}{c|c}
0.4375+U^{\prime}_{1,4} & 0.1875+U^{\prime}_{2,2} \\
0.1875+U^{\prime}_{1,2} & 0.0625+U^{\prime}_{2,1} \\
0.3125+U^{\prime}_{1,3} & 0.8125+U^{\prime}_{2,7} \\
0.8125+U^{\prime}_{1,7} & 0.9375+U^{\prime}_{2,8} \\
0.0625+U^{\prime}_{1,1} & 0.5625+U^{\prime}_{2,5} \\
0.9375+U^{\prime}_{1,8} & 0.6875+U^{\prime}_{2,6} \\
0.6875+U^{\prime}_{1,6} & 0.4375+U^{\prime}_{2,4} \\
0.5625+U^{\prime}_{1,5} & 0.3125+U^{\prime}_{2,3} \\
\end{array}
\right]}^{\mb{Z}_1}
\overbrace{
\left[
\begin{array}{c|c}
0.0625+U^{\prime}_{1,1} & 0.1875+U^{\prime}_{2,2} \\
0.3125+U^{\prime}_{1,3} & 0.0625+U^{\prime}_{2,1} \\
0.9375+U^{\prime}_{1,8} & 0.8125+U^{\prime}_{2,7} \\
0.4375+U^{\prime}_{1,4} & 0.9375+U^{\prime}_{2,8} \\
0.6875+U^{\prime}_{1,6} & 0.5625+U^{\prime}_{2,5} \\
0.1875+U^{\prime}_{1,2} & 0.6875+U^{\prime}_{2,6} \\
0.5625+U^{\prime}_{1,5} & 0.4375+U^{\prime}_{2,4} \\
0.8125+U^{\prime}_{1,7} & 0.3125+U^{\prime}_{2,3} \\
\end{array}
\right]}^{\mb{W}_{-1}}
\end{align}
\endgroup
The designs presented in Eq. (\ref{thirdOracle1}) are then equal to
\begingroup\makeatletter\def\f@size{8.5}\check@mathfonts
\begin{align}
\overbrace{
\left[
\begin{array}{c|c}
0.4375+U^{\prime}_{1,4} & 0.9375+U^{\prime}_{2,8} \\
0.1875+U^{\prime}_{1,2} & 0.6875+U^{\prime}_{2,6} \\
0.3125+U^{\prime}_{1,3} & 0.0625+U^{\prime}_{2,1} \\
0.8125+U^{\prime}_{1,7} & 0.3125+U^{\prime}_{2,3} \\
0.0625+U^{\prime}_{1,1} & 0.1875+U^{\prime}_{2,2} \\
0.9375+U^{\prime}_{1,8} & 0.8125+U^{\prime}_{2,7} \\
0.6875+U^{\prime}_{1,5} & 0.5625+U^{\prime}_{2,5} \\
0.5625+U^{\prime}_{1,5} & 0.4375+U^{\prime}_{2,4} \\
\end{array}
\right]}^{\widetilde{\mb{W}}_{-1}}
\overbrace{
\left[
\begin{array}{c|c}
0.4375+U^{\prime}_{1,4} & 0.1875+U^{\prime}_{2,2} \\
0.1875+U^{\prime}_{1,2} & 0.0625+U^{\prime}_{2,1} \\
0.3125+U^{\prime}_{1,3} & 0.8125+U^{\prime}_{2,7} \\
0.8125+U^{\prime}_{1,7} & 0.9375+U^{\prime}_{2,8} \\
0.0625+U^{\prime}_{1,1} & 0.5625+U^{\prime}_{2,5} \\
0.9375+U^{\prime}_{1,8} & 0.6875+U^{\prime}_{2,6} \\
0.6875+U^{\prime}_{1,6} & 0.4375+U^{\prime}_{2,4} \\
0.5625+U^{\prime}_{1,5} & 0.3125+U^{\prime}_{2,3} \\
\end{array}
\right]}^{\mb{Z}_1}
\overbrace{
\left[
\begin{array}{c|c}
0.0625+U^{\prime}_{1,1} & 0.1875+U^{\prime}_{2,2} \\
0.3125+U^{\prime}_{1,3} & 0.0625+U^{\prime}_{2,1} \\
0.9375+U^{\prime}_{1,8} & 0.8125+U^{\prime}_{2,7} \\
0.4375+U^{\prime}_{1,4} & 0.9375+U^{\prime}_{2,8} \\
0.6875+U^{\prime}_{1,6} & 0.5625+U^{\prime}_{2,5} \\
0.1875+U^{\prime}_{1,2} & 0.6875+U^{\prime}_{2,6} \\
0.5625+U^{\prime}_{1,5} & 0.4375+U^{\prime}_{2,4} \\
0.8125+U^{\prime}_{1,7} & 0.3125+U^{\prime}_{2,3} \\
\end{array}
\right]}^{\mb{W}_{-1}}
\end{align}
\endgroup
\medskip
\begin{Rem}
\label{RemTotatEstim}
The estimator (\ref{TotalIndexEstim}) of the total-order Sobol' index $S_{\{i\}}^{T}$ can be computed with the simulations run at both $\mb{W}_{-i}$ (equal to those run at $\mb{W}$) and $\mb{Z}_i$.
\end{Rem}

\medskip
The Oracle 2 estimator relies on the two designs $\mb{X}$ and $\mb{W}$.
As $\mb{X}$ and $\mb{Z}_i$ are also mutually independent rLHDs, a second Oracle 2 estimator can be computed for each index $S_{\{j\}}$ for $1\leq j\leq d$. Hence, every first-order Sobol' index can be re-estimated by an averaged (double) Oracle 2 estimator using the simulations run at $\mb{X}$, $\mb{W}$ and $\mb{Z}_i$. In addition, point out that the reordering trick applied to $\mb{Z}_i$ can generate a design whose $i$-th column matches that of $\mb{W}_{-i}$, so the index $S_{\{i\}}$ can even be re-estimated by an averaged (triple) Oracle 2 estimator whose accuracy will be compared to that of the averaged (triple) Oracle 1 on a numerical example (see Section \ref{NumericalExamples}). 

\section{The rLHDs-based adaptive strategy}

\label{adaptive_strategy}

\vspace{0.4cm}
According to Section \ref{sec_comp}, the best estimator between Oracle 1 and Oracle 2 for computing first-order Sobol' indices when $X_i$ is free of interaction with the other inputs depends on the magnitude of $S_{\{i\}}$. In real problems where the true values of indices are unknown, the use of rLHDs allows us to compute the whole set of first-order Sobol' indices by Oracle 2 with only $2N$ simulations. Then, the small and moderate Oracle 2 estimates can be computed anew by Oracle 1 to check whether they are accurate or tainted by spurious correlation. Below, we give the details of the two-stage strategy$:$
 
\begin{enumerate}[1)]
\item Estimate the whole set of first-order Sobol' indices by Oracle 2 based on two independent randomized rLHDs $\mb{X}$ and $\mb{W}$. We recommend to take a moderate $N$ between $200$ and $400$ depending on the simulation time of $y(.)$. Bootstrap resampling can be used for deriving confidence intervals for each $\hat{S}_{\{j\}}$ ($1 \leq j\leq d$) \citep{Efron86}.

\medskip
\item Re-compute one at a time the small and moderate Oracle 2 estimates by Oracle 1$:$
\begin{enumerate}[i.]
\item Select the input $X_i$ related to the highest $\hat{S}_{\{i\}}$ among the small and moderate $\hat{S}_{\{j\}}$ computed in the first stage, then run $y(.)$ at a new rLHD $\mb{Z}_i$ (see Section \ref{secOracle1rLHDs}), 
\item Re-compute $\hat{S}_{\{i\}}$ by the averaged (triple) Oracle 1 estimator (\ref{oracle1triple}) based on the simulations run at $\mb{X}$, $\mb{W}$ and $\mb{Z}_i$, 
\item As a bonus, re-compute the estimates $\hat{S}_{\{j\}}$ that have not yet been re-computed by Oracle 1 by the averaged Oracle 2 estimator (see Remark \ref{averagedOracle2} below),
\item Re-compute the bootstrap confidence intervals for each $S_{\{j\}}$ ($1\leq j\leq d$),
\item As a bonus, compute the total-order index estimator $\hat{S}^{T}_{\{i\}}$ by Equation (\ref{TotalIndexEstim}) (see Remark \ref{RemTotatEstim}) along with a bootstrap confidence interval.
\end{enumerate}
\end{enumerate}
The second stage can be repeated as many times as the number of small and moderate Oracle 2 estimates in the first stage. However, the practitioner may exit the second stage without re-estimating all the small and moderate estimates. This can typically happen for Type A models (additive models or close) once the sum of several $\hat{S}_{\{j\}}$ is close enough to $1$ with sufficient accuracy. 

\medskip
The number of simulations spent by the adaptive strategy is equal to $N_{T}:=2N+Nm=N(m+2)$ where $m$ is the number of indices being re-estimated in the second stage. If all the indices are re-estimated ($m=d$), this cost matches that of the Saltelli method to estimate the whole set of first-order and total-order Sobol' indices together in one shot \citep{Saltelli02}.

\begin{Rem}
\label{averagedOracle2}
The $k$-th time the second stage is run, the averaged Oracle 2 estimators can be computed from $k+1$ simple estimators. 
\end{Rem}

\section{Numerical examples}

We begin this section by showing that Oracle 1 is more accurate than Oracle 2 for estimating small and moderate first-order indices in additive models, and conversely for large first-order indices. The Uranie platform was used for this purpose \citep{Blanchard2019}. Then, we will implement the adaptive strategy in three different situations, and analyze the accuracy of estimates.

\label{NumericalExamples}

\subsection{Performance of the estimators}

\begin{figure}[ht!]
\centering
\includegraphics[scale=0.39]
{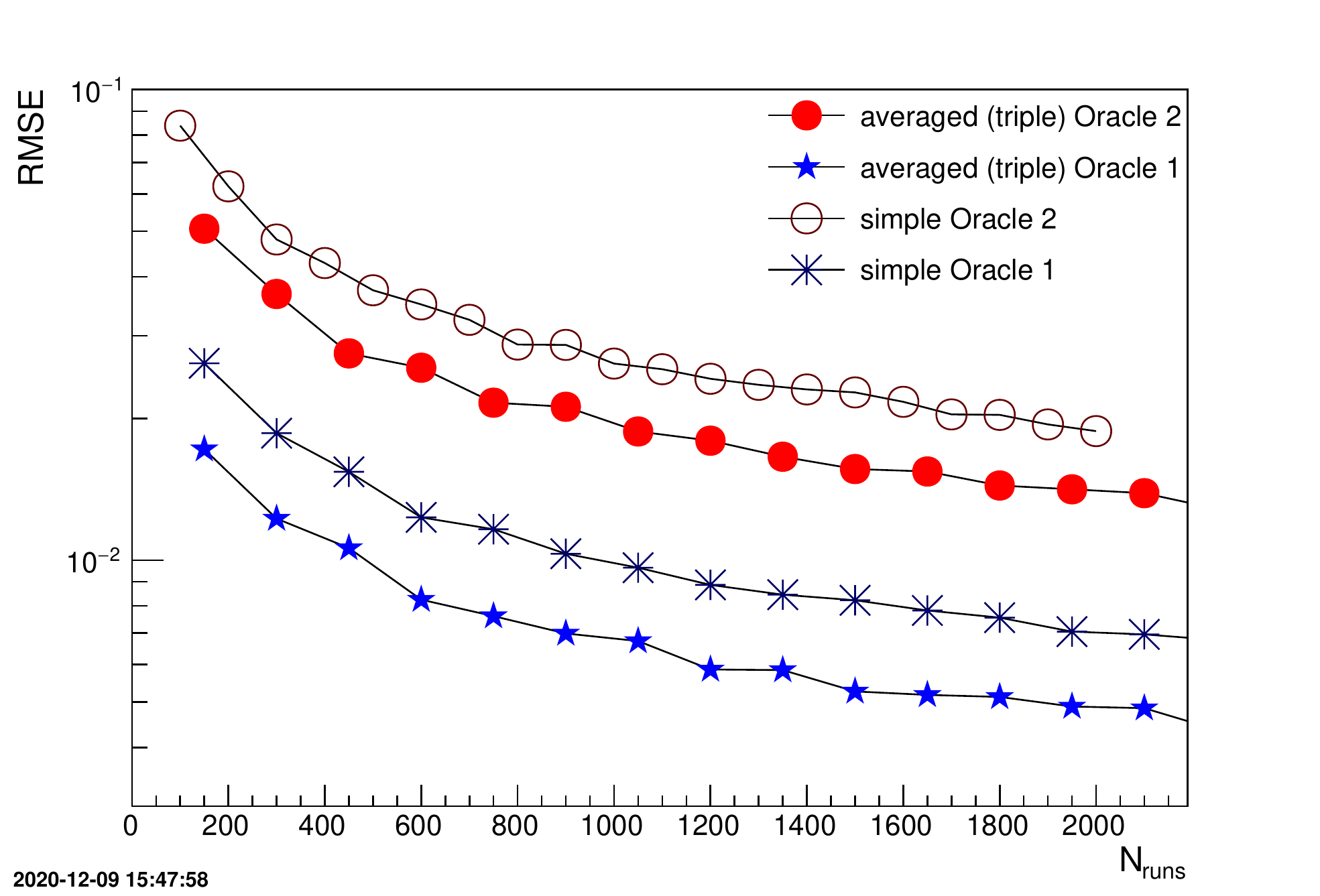}
\hspace{-0.5cm}
\includegraphics[scale=0.39]
{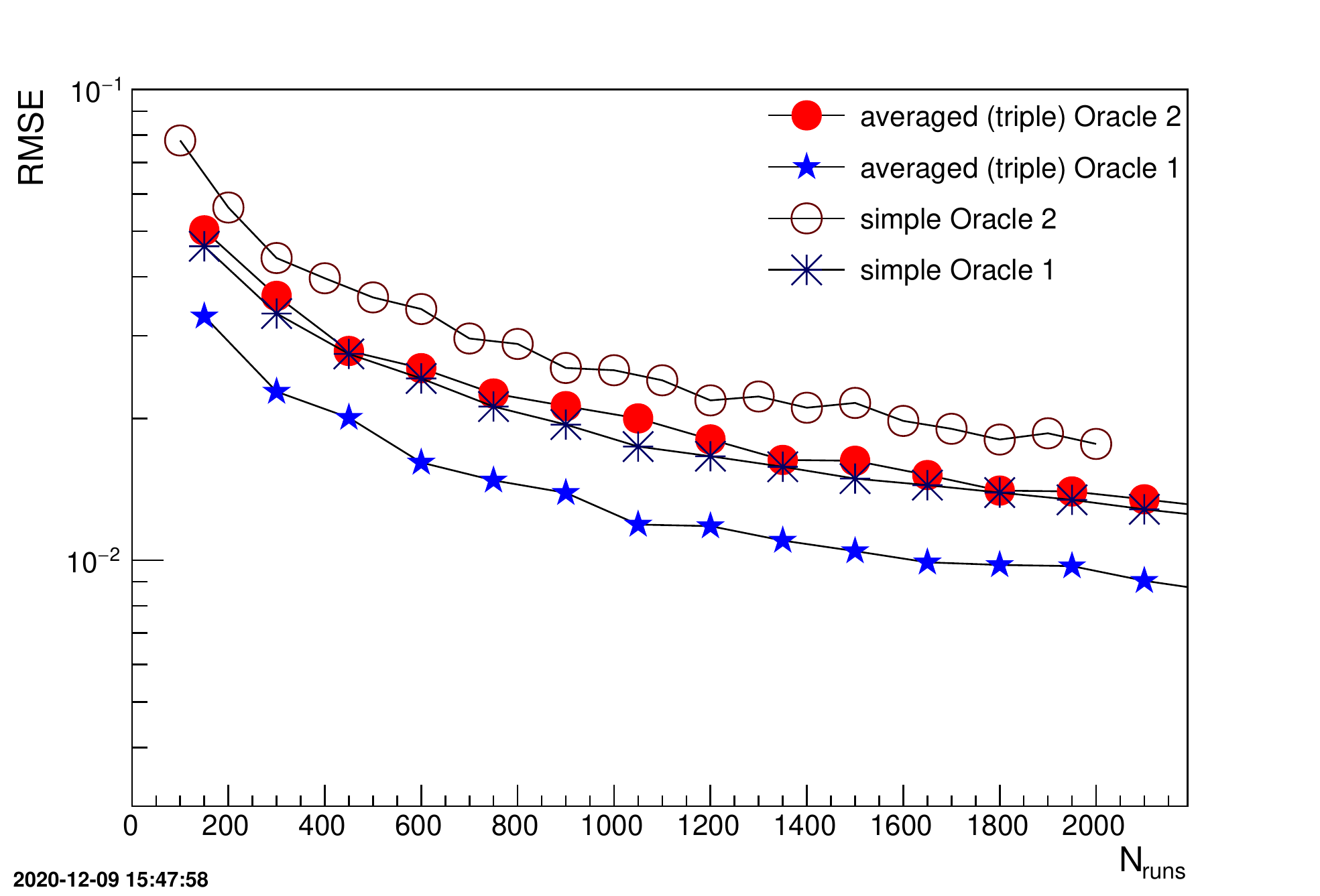}
\includegraphics[scale=0.39]
{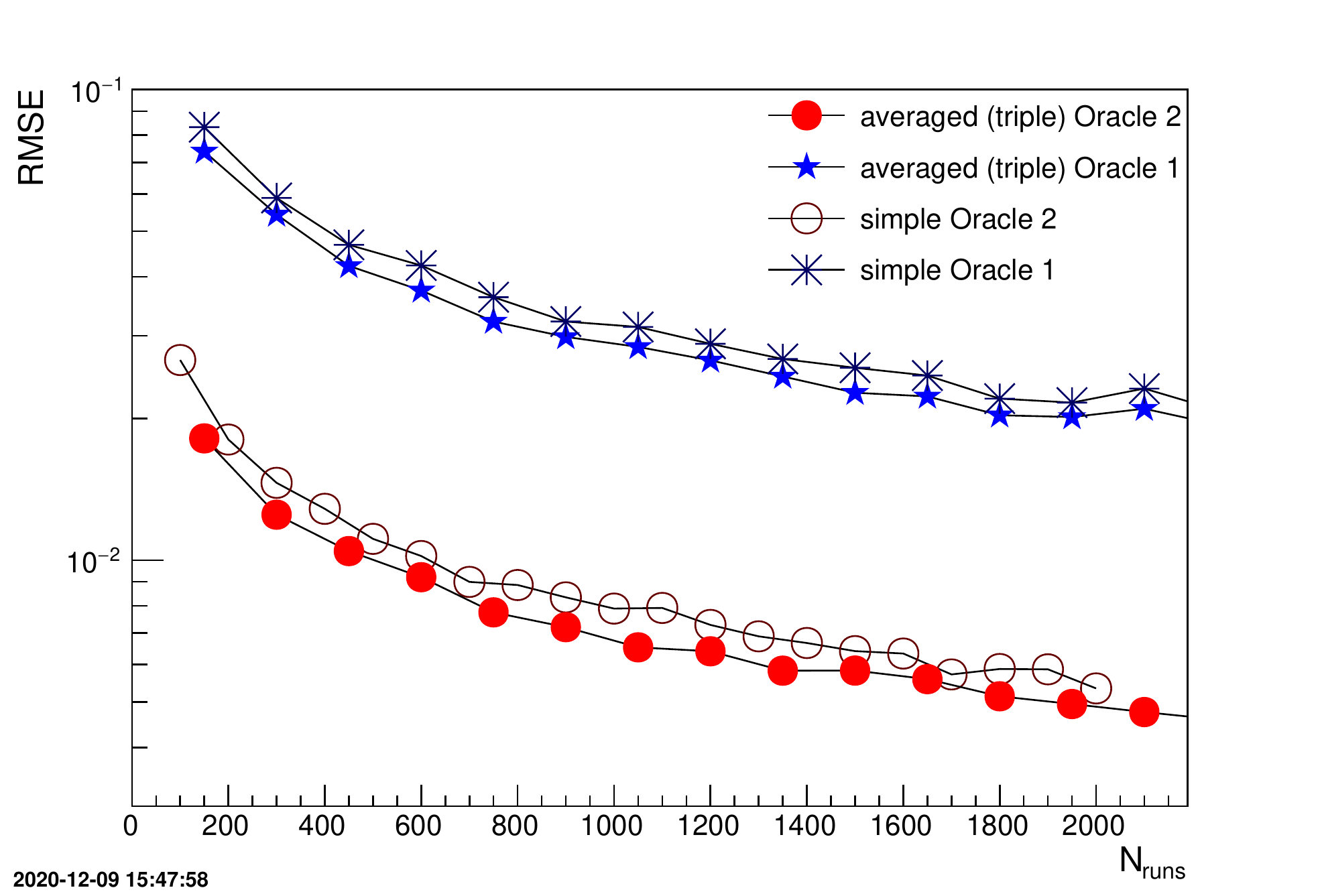}
\caption{Model (\ref{ex2}) : reduction of the RMSE of $\hat{S}_{\{i\}}$ as the number of model evaluations $N_{runs}$ increases$:$ $\hat{S}_{\{1\}}$ (left), $\hat{S}_{\{2\}}$ (right) and $\hat{S}_{\{3\}}$ (bottom). The log scale is used for the $y$-axis.}
\label{figEx2Results}
\end{figure}

\vspace{0.4cm}
The modified Sobol' g-function, used as a benchmark function in \citet{Sob07}, is written as
\begin{equation}
\label{ex2}
y(x_1,x_2,x_3)=\prod_{i=1}^{3}\frac{|4x_i-2|+2+3a_i}{1+a_i},\,\,\,\,\,\,\,\,a_i\neq 1.
\end{equation}
with each $X_i=\mathcal{U}(0,1)$ ($1\leq i\leq d=3$). By setting the coefficients $(a_1,a_2,a_3)$ to $(19,9,4)$, the theoretical values of the first-order Sobol' indices  are equal to
\begin{equation}
S_{\{1\}}=0.0476,\,\,\,\,\,\, S_{\{2\}}=0.1904,\,\,\,\,\,\,S_{\{3\}}=0.7616.
\end{equation}
The performance of the estimators was assessed by the empirical Root Mean Square Error (RMSE) frequently used as a measure of accuracy$:$
\begin{equation}
\label{rmse}
\textrm{RMSE}\,=\,\sqrt{\frac{1}{n}\Big[\sum_{k=1}^{n}\big(\hat{S}_{\{i\}}(k)-S_{\{i\}}\big)^{2}\Big]}.
\end{equation}
For an increasing number of model evaluations $N_{runs}$, the RMSE of each $\hat{S}_{\{i\}}$ ($1\leq i\leq 3$) was evaluated with $n=1000$ replications. Four estimators have been compared with one another$:$
\begin{itemize}
\item the Oracle 2 estimator ($N_{runs}=2N$),
\item the averaged (triple) Oracle 2 estimator ($N_{runs}=3N$),
\item the Oracle 1 estimator ($N_{runs}=3N$),
\item the averaged (triple) Oracle 1 estimator ($N_{runs}=3N$).
\end{itemize}
Fig. \ref{figEx2Results} presents the RMSE reduction as $N_{runs}$ increases. The Oracle 1 estimator (respectively the averaged (triple) Oracle 1 estimator) of $S_{\{1\}}$ outperforms the Oracle 2 estimator (respectively the averaged (triple) Oracle 2 estimator), and conversely for $S_{\{3\}}$. For $S_{\{2\}}$, the averaged (triple) Oracle 2 estimator competes well with the simple Oracle 1 estimator, but is dominated by the averaged version of the latter. The closer the index to $0$, the more the advantage in favor of Oracle 1 is pronounced. This observation meets the theory that when $S_{\{i\}}=S^{T}_{\{i\}}$ the spurious correlation of Oracle 1 increasingly decreases to $0$ as $S_{\{i\}}$ tends toward $0$. Conversely, the spurious correlation of Oracle 2 increasingly decreases to $0$ as $S_{\{i\}}$ tends toward $1$ (see Eqs. (\ref{varO2}) and (\ref{varO1}) in \ref{appendix}). Finally, lower values of RMSE show the benefit of computing the averaged (triple) estimators which can outperform simple estimators based on much more model evaluations.

\subsection{Implementation of the adaptive strategy}

\medskip
In what follows the first two examples implement the two-stage strategy on an additive model where there is a theoretical justification to estimate the small and moderate Sobol' indices by Oracle 1. The third example deals with a complex model including interactions between input variables.

\vspace{0.4cm}
\paragraph{First example} We still consider the modified Sobol' g-function with $(a_1,a_2,a_3)=(19,9,4)$  but adding a linear contribution made up with $7$ extra input variables $X_i\,\, (4\leq i\leq 10)$$:$
\begin{equation}
\label{model_adaptive}
y(\mb{x})=\prod_{i=1}^{3}\frac{|4x_i-2|+2+3a_i}{1+a_i}+\sum_{i=4}^{10} \epsilon_i x_i\,\,\,\,\,\,\,\,a_i\neq 1
\end{equation}

\medskip
The value given to $\epsilon$ tunes the relative variance contribution of each part. By setting $\epsilon_i:=0.10$, then
\begin{equation}
S_{\{1\}}=0.0474,\,\,\,\,\,\, S_{\{2\}}=0.1896,\,\,\,\,\,\,S_{\{3\}}=0.7585,
\end{equation}
and the linear part is negligible$:$
\begin{equation}
S_{\{i\}}=5.9\times 10^{-4}\,\,\,\,\,\,;\,\,\,\,\,\,i=4,\cdots,10.
\end{equation}

\begin{figure}[htbp]
    \hspace*{-2.9em}
    \subfloat[]{
        \includegraphics[width=0.57\textwidth] {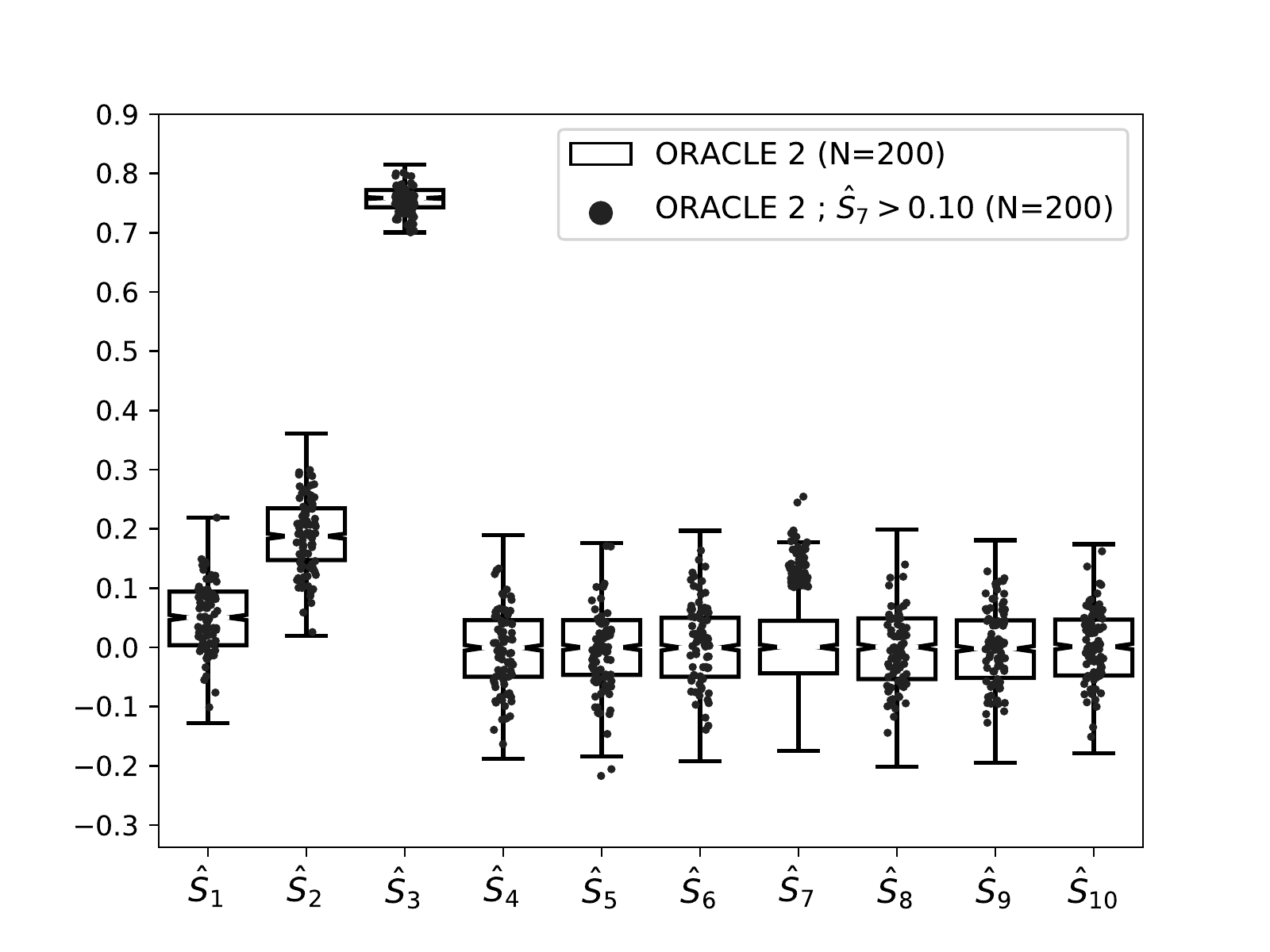}
        \label{fig:img1} } \hspace*{-2.5em}
    \subfloat[]{
        \includegraphics[width=0.57\textwidth] {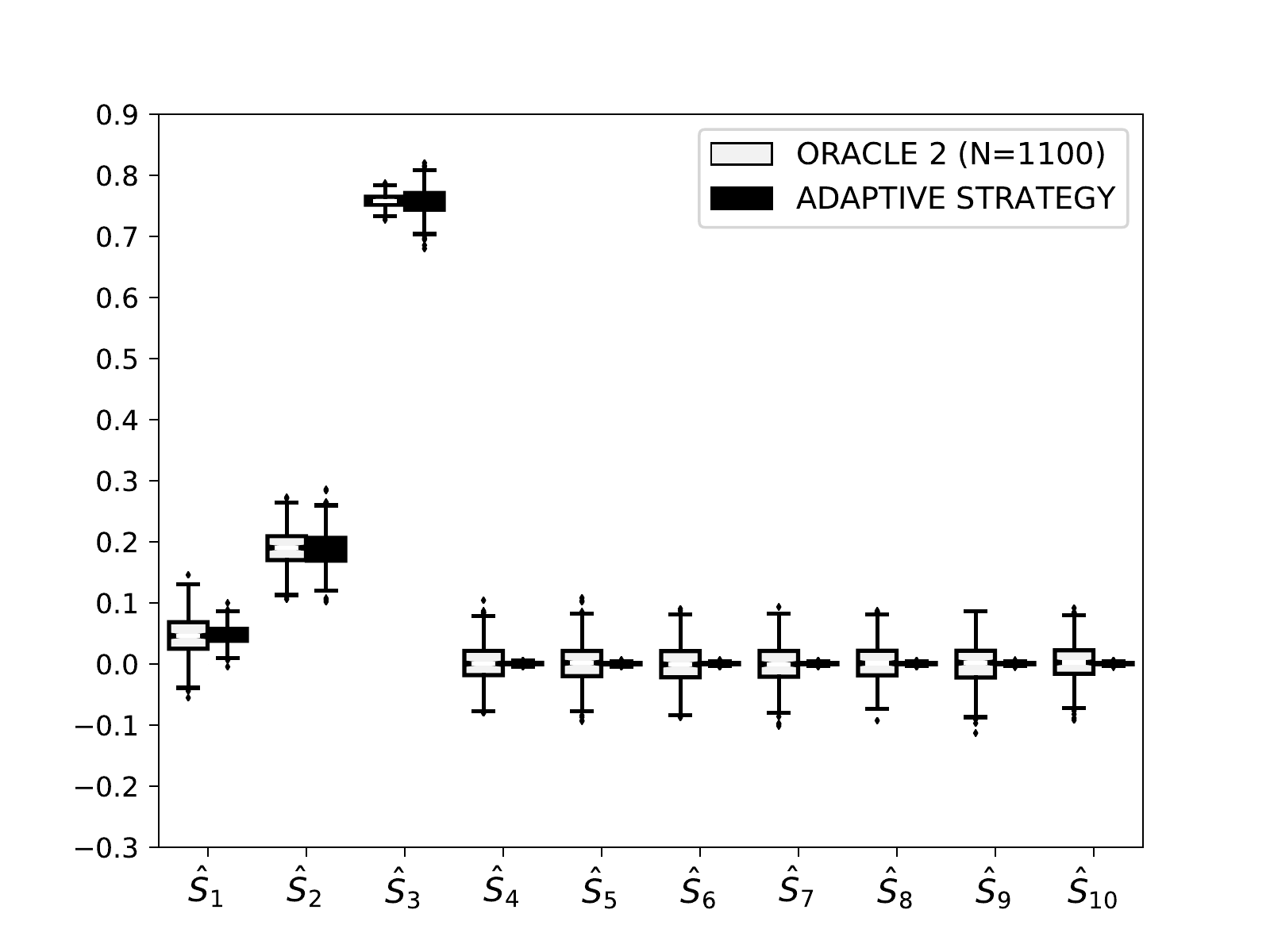}
        \label{fig:img2} } \\
    \caption{(a) Box plots of one thousand estimates $\hat{S}\in \rr^{10}$ computed by Oracle 2 with $N=200$. Those where $\hat{S}_7>0.10$ are marked by black dots; (b) Box plot of one thousand estimates computed by Oracle 2 with $N=1100$ (white boxes) compared to those deriving from the adaptive strategy starting from $N=200$, then where all the small and moderate indices were recomputed by the averaged (triple) Oracle 1 (black boxes).}
    \label{pcp_adaptive}
\end{figure}

\medskip
Model (\ref{model_adaptive}) has one large first-order effect as well as two moderate ones. Fig. \ref{pcp_adaptive}-(a) shows box plots consisting of one thousand independent estimates $\hat{S}:=(\hat{S}_1,\hat{S}_2,\cdots,\hat{S}_{10})$ with each $\hat{S}_{\{i\}}\,\,(1\leq i\leq 10)$ being computed by Oracle 2 with $N=200$. As expected, the large first-order effect related to $X_3$ is accurately estimated by Oracle 2 whereas every other index is indistinguishably either accurately estimated or tainted by spurious correlation. We indeed calculated $40.4\%$ of estimates $\hat{S}$ where at least one $\hat{S}_{\{i\}}\,\,(4\leq i\leq 10)$ was above $0.10$. Such a spurious correlation is related to the large variance of Oracle 2. As an example of that, the black dots in Fig. \ref{pcp_adaptive}-(a) refer to the subset of estimates where $\hat{S}_{\{7\}}$ is above $0.10$. By contrast, we know that $S_{\{7\}}$ as well as the other six negligible indices will be properly estimated by the averaged (triple) Oracle 1 estimators. The adaptive strategy has thus been applied to every $\hat{S}$. Fig. \ref{pcp_adaptive}-(b) shows the new estimates after re-computing the nine possible small and moderate Sobol' indices by the averaged (triple) Oracle 1. This required a total of $N_{T}=2200$ simulations ($N_T=2N+9N$ with $N=200$). On the same figure, we displayed one thousand of independent Oracle 2 estimates computed with $N_{T}=2N=2\times 1100$ simulations. We can see a strong advantage to the two-stage strategy in terms of accuracy of estimates since the risk of inacurrate small indices has fully vanished. 

\medskip
The second stage of the proposed strategy re-computes the small and moderate first-order Oracle 2 estimates by starting from the highest ones. Note that the small indices can also be wrongly estimated by Oracle 2 in case of negative spurious correlation (causing negative estimates). Actually in this first example, $\hat{S}_{\{1\}}$ was sometimes re-estimated at a late time in the second stage.

\begin{figure}[htbp]
    \hspace*{-2.9em}
    \subfloat[]{
        \includegraphics[width=0.57\textwidth] {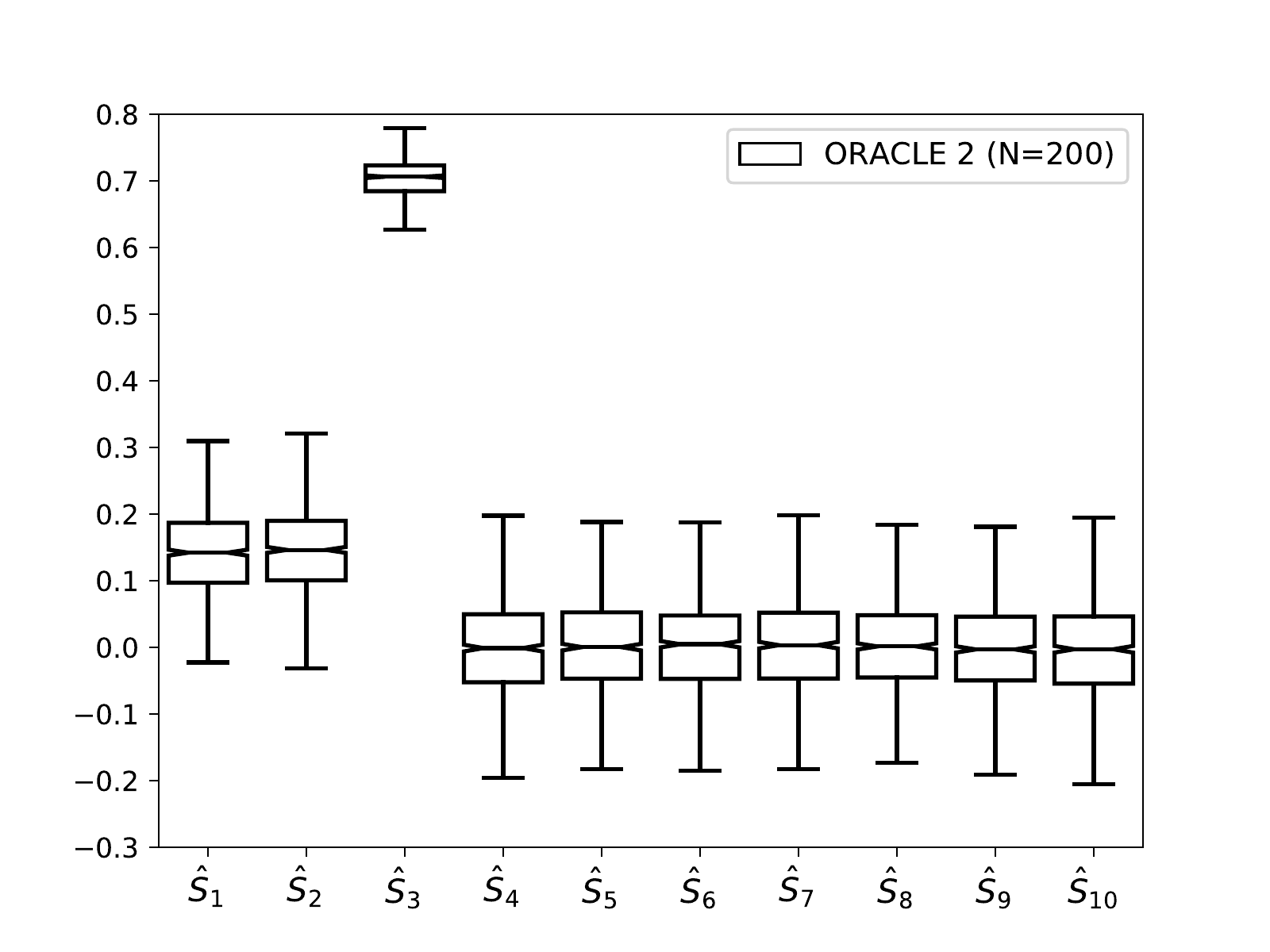}
        \label{fig:img1} } \hspace*{-2.5em}
    \subfloat[]{
        \includegraphics[width=0.57\textwidth] {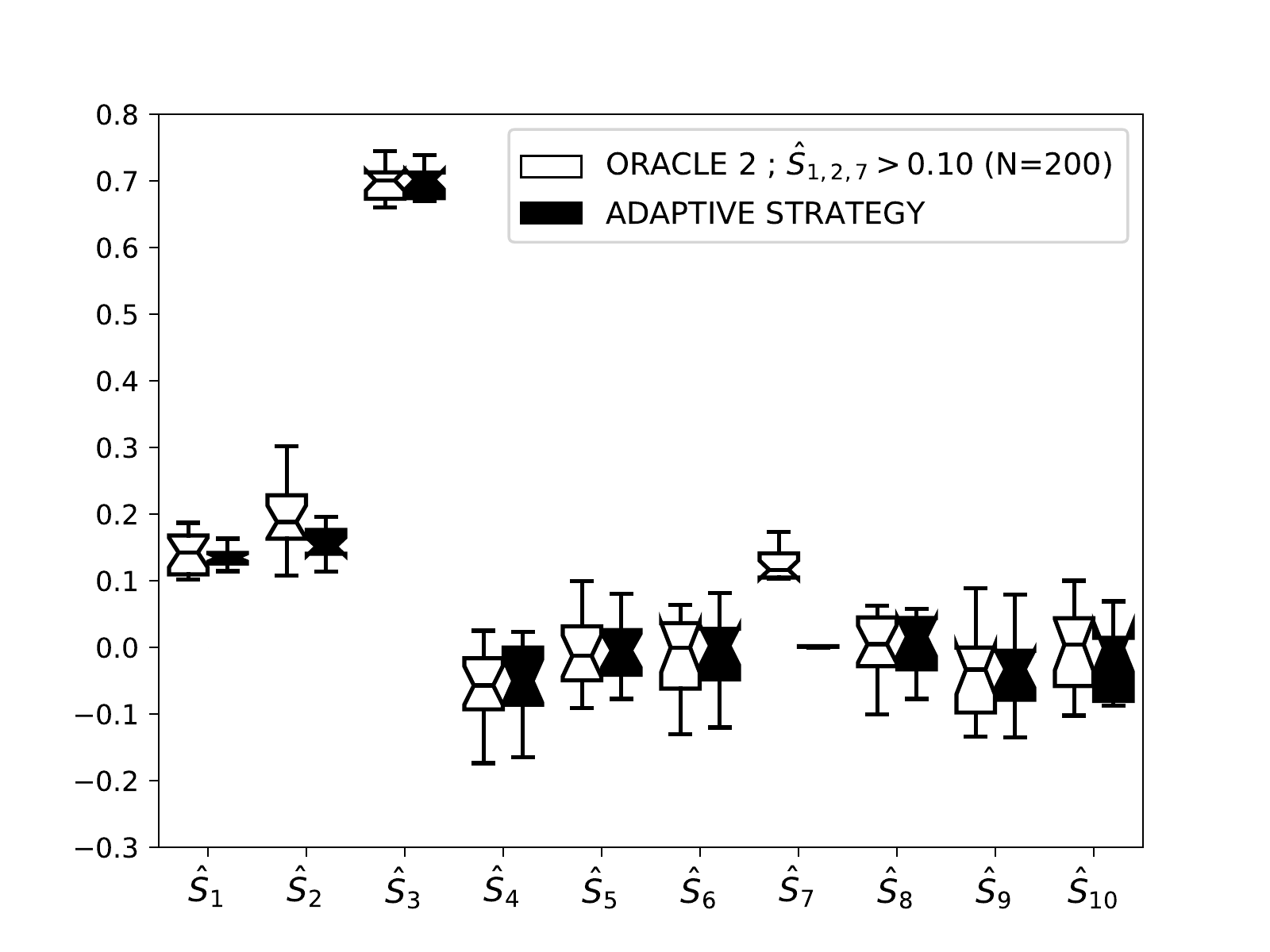}
        \label{fig:img2} } \\
    \caption{(a) Box plots of one thousand estimates $\hat{S}\in \rr^{10}$ computed by Oracle 2 with $N=200$. (b) Box plots of those where $\hat{S}_{\{1\}}$, $\hat{S}_{\{2\}}$ and $\hat{S}_{\{7\}}$ are above $0.10$ (white boxes) and the corresponding estimates after re-computing them by the second stage of the adaptive strategy (black boxes).}
    \label{pcp_adaptive_2}
\end{figure}

\medskip
\paragraph*{Second example} In some cases, re-estimating all the possible small and moderate indices may not be necessary. Let us now consider Model (\ref{model_adaptive}) with $(a_1,a_2,a_3)$ being equal to $(10,10,4)$. The model is still additive such that
\begin{equation}
S_{\{1\}}=0.1456,\,\,\,\,\,\, S_{\{2\}}=0.1456,\,\,\,\,\,\,S_{\{3\}}=0.7046,
\end{equation}
and
\begin{equation}
S_{\{i\}}=5.4\times 10^{-4}\,\,\,\,\,\,;\,\,\,\,\,\,i=4,\cdots,10.
\end{equation} 
Fig. \ref{pcp_adaptive_2}-(a) shows one thousand estimates $\hat{S}$ from the first stage with $N=200$. The white boxes in Fig. \ref{pcp_adaptive_2}-(b) show those where $\hat{S}_{\{1\}}$, $\hat{S}_{\{2\}}$ and $\hat{S}_{\{7\}}$ are above $0.10$. The second stage of the adaptive strategy was then run three times to re-compute these estimates by the averaged (triple) Oracle 1 estimator. The updated estimates, displayed by black boxes on Fig. \ref{pcp_adaptive_2}-(b), show that $S_{\{7\}}$ is in fact close to $0$ which allows us to conclude that the variance of $Y$ is mostly explained by $X_1$, $X_2$ and $X_3$ in an additive way. At this point the practitioner could then decide to exit the second stage if he considers that it is useless to accurately estimate the remaining variables (since having very small or negligible impact).

\begin{figure}[ht!]
    \hspace*{-2.9em}
    \subfloat[]{
        \includegraphics[width=0.58\textwidth] {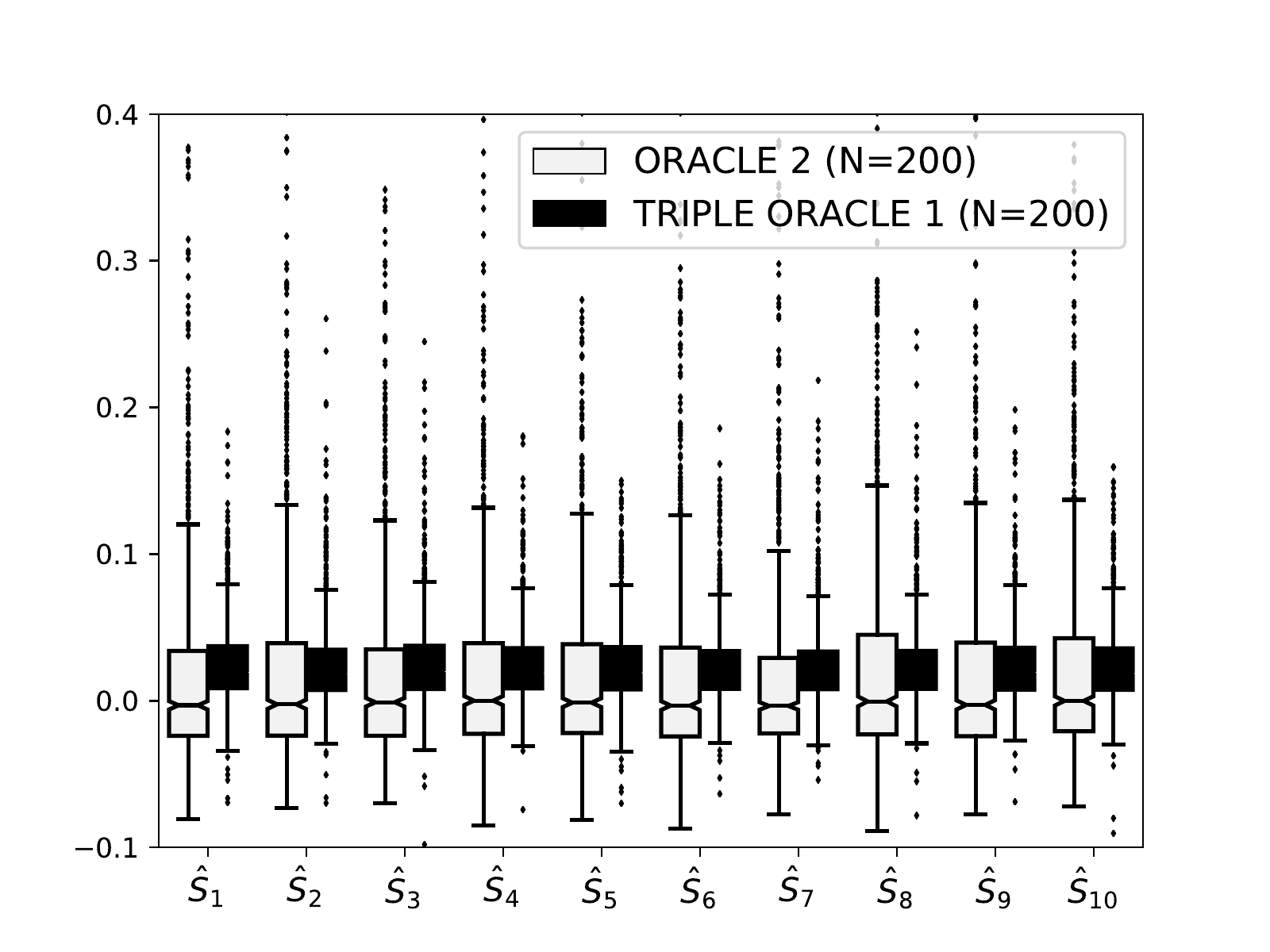}
        \label{fig:img1} } \hspace*{-3.0em}
    \subfloat[]{
        \includegraphics[width=0.58\textwidth] {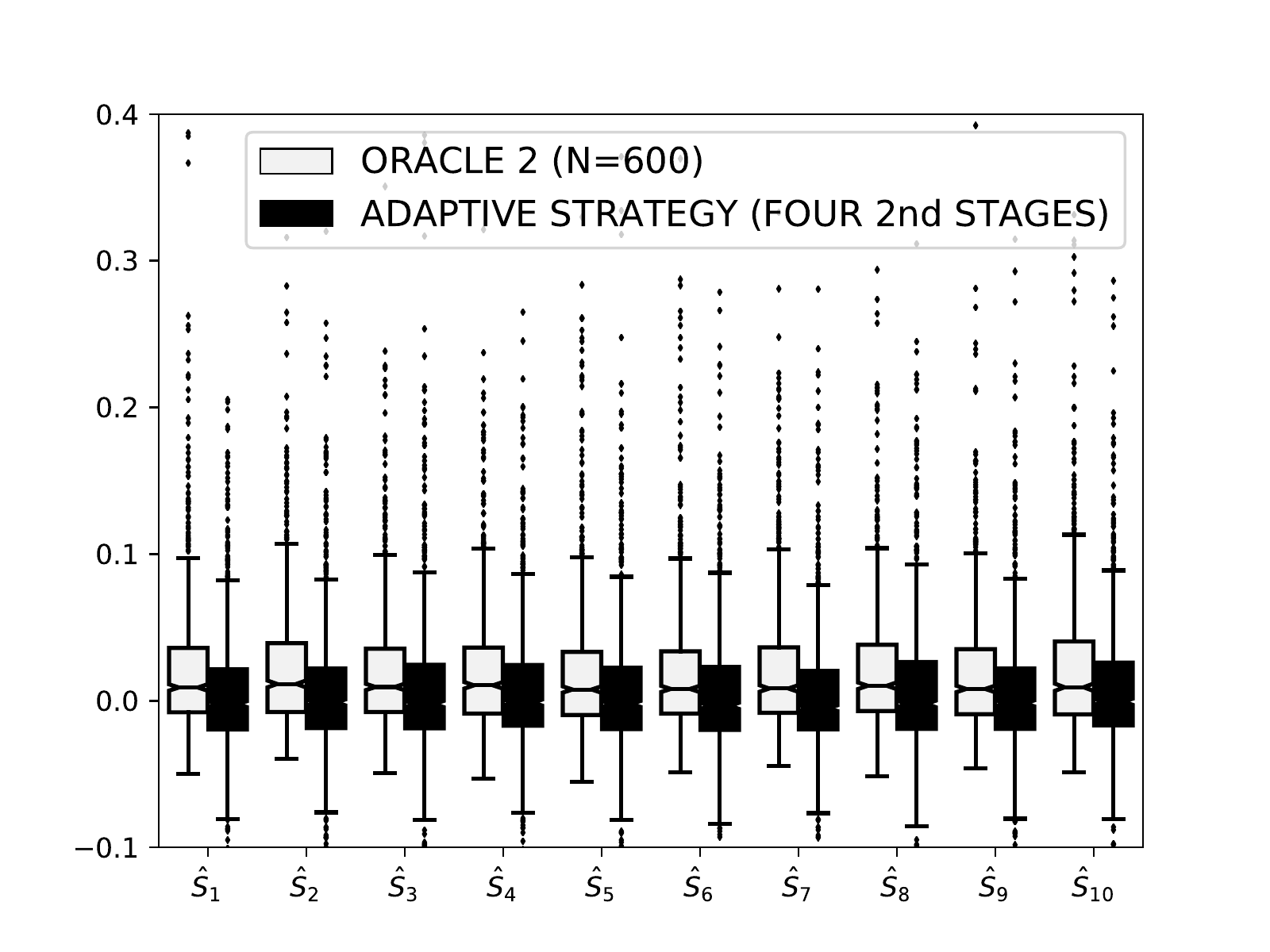}
        \label{fig:img2} } \\
    \caption{(a) Box plots of one thousand estimates $\hat{S}\in \rr^{10}$ computed by Oracle 2 with $N=200$ (white boxes) and by the averaged (triple) Oracle 1 (black boxes); (b) Box plots of one thousand estimates $\hat{S}\in \rr^{10}$ computed by Oracle 2 with $N=600$ (white boxes) against those derived from the adaptive strategy starting from $N=200$, then followed by four loops of the second stage (black boxes).}
    \label{pcp_adaptive_3}
\end{figure}

\medskip
\paragraph*{Third example} We are also interested in non-additive models including interactions between input variables.  According to Section \ref{sec_comp}, there is no rule to choose between Oracle 1 and Oracle 2 when $S_{\{i\}}^{T}>>S_{\{i\}}$. Being able to estimate the total Sobol' indices in the second stage, the adaptive strategy remains well-suited. We focused on the standard g-Sobol function, written as 
\begin{equation}
\label{model_gSobol}
y(\mb{x})=\prod_{i=1}^{10}\frac{|4x_i-2|+a_i}{1+a_i}
\end{equation}
with $a_i=0$ $(1\leq i\leq 10)$. This is a Type C model whose Sobol' indices can be calculated analytically \citep{Sob07}$:$
\begin{equation}
S_{\{i\}}=1.989\times 10^{-2}\,\,\,\,\,\,;\,\,\,\,\,\,i=1,\cdots,10
\end{equation}
and 
\begin{equation}
S^{T}_{\{i\}}=0.8210\,\,\,\,\,\,;\,\,\,\,\,\,i=1,\cdots,10.
\end{equation}
Fig. \ref{pcp_adaptive_3}-(a) still presents one thousand independent estimates $\hat{S}$ computed with $N=200$ respectively by Oracle 2 (white boxes) and the averaged (triple) Oracle 1 estimator (black boxes). The latter outperforms Oracle 2, which shows a possible interest for the adaptive strategy. Then, we implemented$:$
\begin{itemize}
\item the Oracle 2-based one-shot strategy with $N=600$,
\item the adaptive strategy where the four highest Oracle 2 estimates were re-computed by the averaged (triple) Oracle 1 estimator after running the first stage with $N=200$.
\end{itemize}
The total number of simulations is $N_{T}=1200$ in either case. One thousand estimates $\hat{S}$ were computed in each strategy. These are plotted in Fig. \ref{pcp_adaptive_3}-(b) and Table \ref{ex3RMSE} gives the corresponding RMSE values for each index $\hat{S}_{\{i\}}$ ($1\leq i \leq 10$). Overall, we can conclude that the two strategies are comparable to each other.

\medskip
In Section \ref{applicationNuclear} hereafter, we apply the adaptive strategy to a computer model from the nuclear field.

\begin{table}
\begingroup\makeatletter\def\f@size{6.5}\check@mathfonts
\setlength\extrarowheight{2pt}
\vspace{0.3cm}
\centering
\begin{tabular}{|l|c|c|c|c|c|c|c|c|c|c|}
\hline
  & $\hat{S}_1$ & $\hat{S}_2$ & $\hat{S}_3$ & $\hat{S}_4$ & $\hat{S}_5$ & $\hat{S}_6$ & $\hat{S}_7$ & $\hat{S}_8$ & $\hat{S}_9$ & $\hat{S}_{10}$ \\
\hline
\scriptsize{RMSE of the one-shot strat.} & $0.050$ & $\mb{0.049}$ & $\mb{0.046}$ & $\mb{0.045}$ & $\mb{0.048}$ & $0.051$ & $0.053$ & $\mb{0.046}$ & $0.052$ & $0.051$\\
\hline 
\scriptsize{RMSE of the adaptive strat.} & $\mb{0.047}$ & $0.053$ & $0.052$ & $0.051$ & $0.051$ & $\mb{0.050}$ & $\mb{0.050}$ & $0.057$ & $\mb{0.051}$ & $\mb{0.049}$\\
\hline
\end{tabular}
\caption{RMSE values corresponding to the estimates displayed in Fig. \ref{pcp_adaptive_3}-(b). Those printed in bold are the smallest in pairs.}
\label{ex3RMSE}
\endgroup
\end{table}

\section{Application in the nuclear field}
\label{applicationNuclear}

Nuclear research reactors are of strategic importance to support commercial nuclear power plants, develop new technologies for future reactors, and produce radioisotopes for research and medical applications \citep{Ghione2017}. In the core region of these reactors, the coolant usually flows at relatively low pressure ($<$1 Mpa) in narrow channels that allow high-performance heat removal capabilities within compact volumes. These channels are arranged in a parallel configuration and no cross flow occurs. Such an arrangement may be subject to the so-called flow excursion instability \citep{Ledi1938}. In fact, uneven distributions of power and flow over the core may lead to flow starvation and eventually boiling crisis in some of the channels. This instability is a primary concern in research reactors operating at low pressure due to larger vapor-to-liquid density ratio. Reliable and precise simulations of this phenomenon are therefore essential. The thermal-hydraulic system code CATHARE is used for safety analysis studies \citep{Geffraye2011}. This code was previously validated against flow excursion experiments proving good performances \citep{Ghione2017_2}. 

In order to obtain a better understanding of the code behavior when simulating the flow excursion instability and determining the most influential parameters, a sensitivity analysis is carried out. The study focuses on the simulation of a Whittle-Forgan flow excursion experiment \citep{Whittle1967}, performed in a uniformly heated vertical narrow rectangular channel with gap size of 3.23 mm and upward flow. In the experiment, the outlet pressure (0.12 MPa), inlet temperature ($55^{\circ}$C) and heat flux (1.04 $MW/m^2$) were fixed, while the mass flux was decreased in steps until the Onset of Flow Instability (OFI) could be identified. The experimental mass flux at OFI (i.e. the Quantity of Interest) was therefore found equal to 2356.5 $kg/m^{2}/s$. The code CATHARE can reproduce the flow excursion phenomenon in a fully satisfactory way and a relative error of less than 0.6 \% is obtained on the QoI \citep{Ghione2020}.

\medskip

The list of uncertainty sources along with the associated probability distributions, assumed independent one another, is compiled in Table \ref{SelectedInputs}. The uncertainties on the geometry, the initial and boundary conditions have been determined via expert judgment while those corresponding to the CATHARE closure laws rely on a literature review of previous works at CEA. The latter are applied to the closure laws through multiplicative factors. For example, the single-phase friction factor f is modified with the multiplicative parameter SP1CL as:
\begin{equation}
f=SP1CL\times f.
\end{equation}

\begin{table}[ht!]
\begingroup\makeatletter\def\f@size{8.5}\check@mathfonts
\setlength\extrarowheight{2pt}

\vspace{0.3cm}
\centering
\begin{tabular}{|l|c|c|}
\hline
\textbf{Input parameter} & \textbf{Distribution}  & \textbf{Range} \\
\hline
\multicolumn{3}{|c|}{CATHARE closure laws} \\
\hline
1: Subcooled condensation (SP1QLE) & Log-uniform  &  $[0.3,3.0]$ \\
\hline
2: Single-Phase friction factor (SP1CL) & Normal & $[0.92,1.08]$ \\
\hline
3: Two-phase friction mutliplier (P1CLGN) &  Uniform & $[0.8,1.2]$ \\
\hline
4: Interfacial friction (SP1TOI) &  Log-normal & $[0.4,2.2]$ \\
\hline
5: NVG point (P1NVGP) & Normal & $[0.85,1.15]$ \\
\hline
6: Wall heat transfer in nucleate boiling (PCNB) & Normal & $[0.56,1.44]$ \\
\hline
7: Wall heat transfer in turbulent forced convection (PCFLT) & Log-normal & $[0.5,2.0]$ \\
\hline
8: Wall heat transfer in laminar forced convection (PCFLL) & Log-normal & 
$[0.5,2.0]$ \\
\hline
9: Wall heat transfer in turbulent natural convection (PCNLT) & Log-normal & 
$[0.5,2.0]$ \\
\hline
10: Wall heat transfer in laminar natural convection (PCNLL) & Log-normal & 
$[0.5,2.0]$ \\
\hline
\multicolumn{3}{|c|}{Geometry, initial and boundary conditions} \\
\hline
11: Gap & Uniform & $\pm 10\%$ \\
\hline
12: Inlet temperature & Uniform & 
$\pm 1^{\circ}C $\\
\hline
13: Heat flux & Uniform & 
$\pm 1.5\%$ \\
\hline
14: Oulet pressure & Uniform & 
$\pm 1.0\%$ \\
\hline
\end{tabular}
\caption{Selected inputs uncertainties: ranges and probability distributions \citep{Ghione2020}}

\label{SelectedInputs}
\endgroup
\end{table}

\begin{figure}[ht!]
\centering
\includegraphics[scale=0.70]
{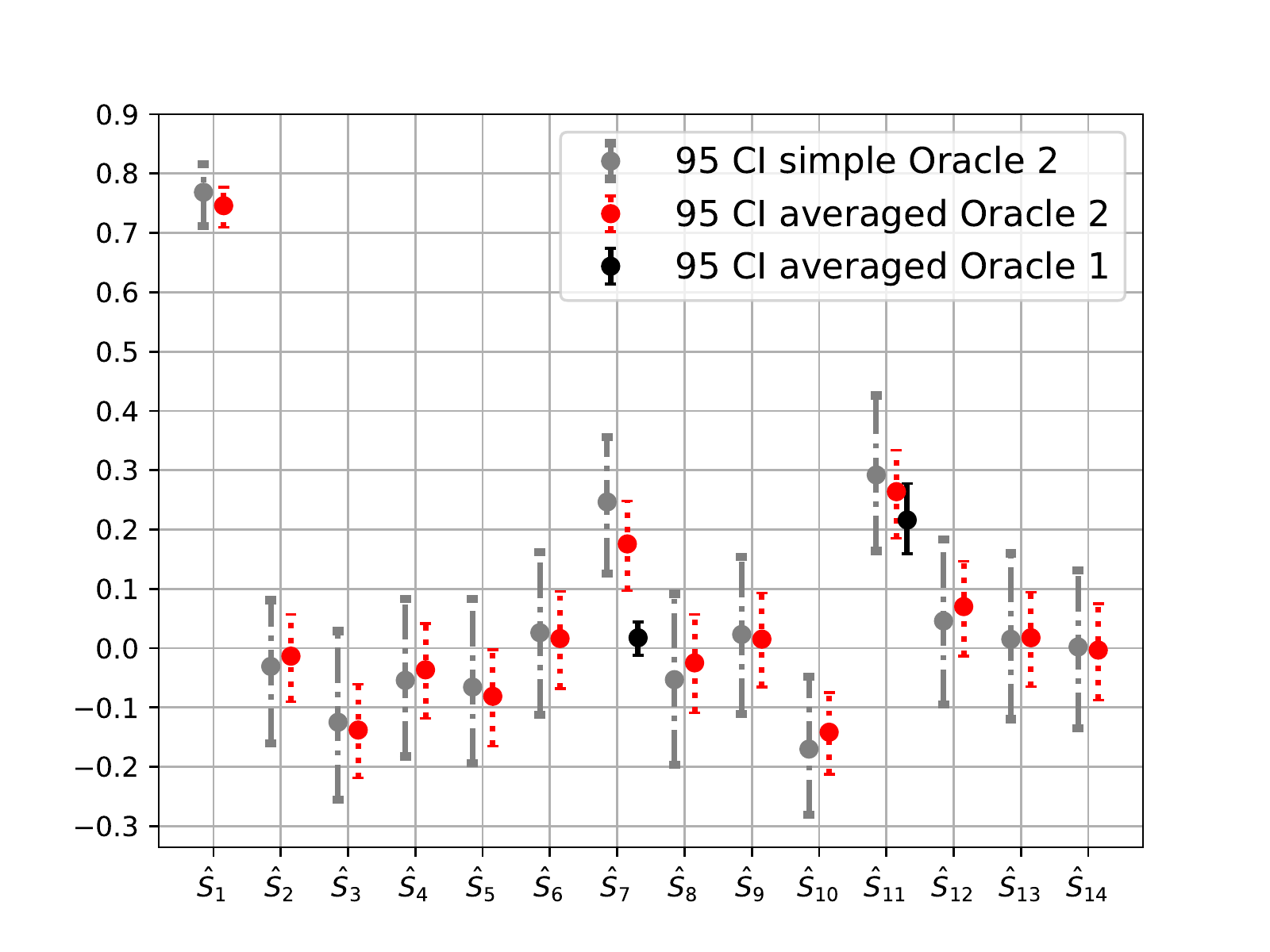}
\vspace{-0.4cm}
\caption{The estimates of the first-order Sobol' indices by the two-stage strategy for the nuclear application}
\label{ResultsNuclearCase}
\end{figure}

\medskip
Two rLHDs $\mb{X}$ and $\mb{W}$ made up with $200$ input locations were sampled according to the probability distributions in Table \ref{SelectedInputs}. We ran the corresponding CATHARE simulations, then retrieved the OFI output values for computing the Oracle 2 estimators of every first-order Sobol' index along with a bootstrap confidence interval. The results show the strong impact of the subcooled condensation as well as two moderate effects related to the wall heat transfer in turbulent forced convection (PCFLT) and the gap$:$
\begin{equation}
\hat{S}_{\{1\}}=0.77,\,\,\,\,\,\,\,\,\,\hat{S}_{\{7\}}=0.25,
\,\,\,\,\,\,\,\,\,\hat{S}_{\{11\}}=0.29.
\end{equation}
Surprisingly, however, the sum of these three indices greatly exceeds $1$, which looks inconsistent. As Oracle 2 can poorly estimate small and moderate indices in presence of spurious correlation, both $S_{\{11\}}$ and $S_{\{7\}}$ were re-estimated by the averaged (triple) Oracle 1 estimator (\ref{oracle1triple}). This required two extra sets of $200$ simulations for each (see Section \ref{secOracle1rLHDs}). The revised estimates are equal to
\begin{equation}
\hat{S}_{\{7\}}=0.018,
\,\,\,\,\,\,\,\,\,\hat{S}_{\{11\}}=0.22,
\end{equation}
showing that PCFLT is not influential, actually. The corresponding $95\%$-confidence bootstrap intervals are reported in Table \ref{FirstIndices7_11}. We also re-estimated all the other indices by the averaged Oracle 2 estimators (Step iii. of the second stage of the adaptive strategy), then updated the confidence intervals. The sizes of these intervals have been significantly reduced. Fig. \ref{ResultsNuclearCase} highlights every $\hat{S}_{\{i\}}$ along with a bootstrap confidence interval $(1\leq i\leq 14)$. In the first stage, the wrong estimation of $S_{\{7\}}$ is due to a strong spurious correlation between the first columns of $\mb{X}$ and $\mb{W}_{-7}$ which is then propagated to the Oracle 2 estimator in proportion to the large $S_{\{1\}}$. This happened because the OFI is observed as quite linear in PCFLT. The re-estimation by Oracle 1 enabled to fix it.

Furthermore, the total-order Sobol' indices of PCFLT and the gap were computed. Table \ref{TotalIndices7_11} shows that $\hat{S}^{T}_{\{7\}}$ is small but non-zero while $\hat{S}^{T}_{\{11\}}$ is close to $\hat{S}_{\{11\}}$. Thus, PCFLT slightly interacts with some other variables.

\begin{table}
\setlength\extrarowheight{2pt}
\vspace{0.3cm}
\centering
\begin{tabular}{|l|l|l|c|}
\hline
Input parameter & 95 CI lower & $\hat{S}$ & 95 CI upper \\
\hline
7: PCFLT & $-0.012$ & $0.018$ & $0.045$ \\
\hline 
11: Gap & $0.160$ &  $0.216$ & $0.278$ \\
\hline
\end{tabular}
\caption{First-order Sobol' indices of both PCFLT and the gap estimated by the averaged (triple) Oracle 1 estimator}
\label{FirstIndices7_11}
\end{table}

Fig. \ref{ResultsNuclearCase} shows (near) negative confidence intervals related respectively to $X_{10}$ and $X_3$. This is explained by a strong negative spurious correlation in $\hat{S}_{\{10\}}$ and $\hat{S}_{\{3\}}$. However, the sum of $\hat{S}^{T}_{\{7\}}$, $\hat{S}_{\{1\}}$ and $\hat{S}_{\{11\}}$ is about equal to $1$, which allowed us to conclude that $S_{\{i\}}\approx 0$ if $i\neq 1,11$. 
 
Finally, the most influential parameter is thus SP1QLE, which modifies the sub-cooled condensation model in CATHARE. A moderate effect of the gap size between the heating plates is observed as well. The sub-cooled condensation affects the formation of void fraction in the channel. Enhancing the condensation leads to a reduction of the void fraction, which reduces the pressure drop in the channel and therefore delays the occurrence of OFI. Analogously, the gap size influences the pressure drop in the channel, and consequently the OFI point.

\begin{table}
\setlength\extrarowheight{2pt}
\vspace{0.3cm}
\centering
\begin{tabular}{|l|l|l|c|}
\hline
Input parameter & 95 CI lower & $\hat{S}^{T}$ & 95 CI upper \\
\hline
7: PCFLT & $0.023$ & $0.052$ & $0.103$ \\
\hline 
11: Gap & $0.168$ &  $0.221$ & $0.292$ \\
\hline
\end{tabular}
\caption{Total-order Sobol' indices of both PCFLT and the gap}
\label{TotalIndices7_11}
\end{table}

\section{Conclusions}

Sobol' indices are largely used for conducting sensitivity analyses of "black-box" computer models. 
The present work has primarily focused on two particular estimators of first-order Sobol' indices, called Oracle 1 and Oracle 2 and was inspired by the following papers from the state of the art$:$ 
\begin{itemize}
\item \cite{Owen13} which derives the variance of four classes of estimators for first-order Sobol' indices, including Oracle 1 and Oracle 2;
\item \cite{Saltelli02} which shows that $N(d+2)$ simulations are enough to compute the whole set of first-order and total Sobol' indices by Oracle 2; 
\item \cite{Glen2012} where the accuracy of Oracle 1 and Oracle 2 are compared to each other in terms of spurious correlation. The larger the variance is, the stronger the spurious correlation is likely to be;
\item \cite{Tissot15} which presents both the practical and theoretical relevance of rLHDs-based Oracle 2 estimators to compute the whole set of first-order Sobol' indices using only $2N$ simulations.
\end{itemize}

\medskip
On the one hand, we have extended the use of rLHDs to the Oracle 1 class of estimators. This class theoretically outperforms Oracle 2 for small and moderate indices, provided that the corresponding input variables are free of interactions. Although being of no help to reduce the computation cost of Oracle 1, the permutation structure of rLHDs can nevertheless provide an averaged (triple) version of this estimator which is more accurate than the simple one.

\medskip
On the other hand, we have proposed a strategy for computing Sobol' indices to take advantage of rLHDs while guaranteeing the accuracy of estimates of small and moderate first-order indices. The strategy starts by computing all the first-order indices by Oracle 2 so that the possible dominant first-order effect(s) can be screened. Then, the small and moderate indices can be sequentially re-estimated by the averaged (triple) Oracle 1 estimator to achieve better accuracy. If Oracle 2 does not show up any large first-order effect in the first stage, then either all the first-order Sobol' indices are small and moderate, or the model is made up of significant interactions between variables (Type B or C models). Although in the latter case there is no theoretical justification for computing small and moderate indices by Oracle 1, the adaptive strategy can still be relevant at least because the total Sobol' indices are estimated in the second stage. The cost of the adaptive strategy is bounded by $N(d+2)$ simulations, which is that of the Saltelli method.

\medskip
The numerical examples have shown that the adaptive strategy is highly cost-saving for additive models or close (Type A models). We have also applied the adaptive strategy to a Type C model where the estimates of the first-order Sobol' indices have been comparable to those computed by Oracle 2 in one shot. Finally, the two-stage strategy was effective in carrying out the sensitivity analysis of a computer model from the nuclear field. 

\medskip
The paper has focused on comparisons between Oracle 1 and Oracle 2 estimators, however, Owen also studied two other classes of estimators, called Correlation 1 and Correlation 2 \citep{Owen13}. Two rules were reported by Owen about these estimators. The first one is that Correlation 1 is less accurate than Oracle 1 when the mean of the model output is large. The second one is that Correlation 2 is extremely accurate for estimating small first-order Sobol' indices because its variance is not at all affected by the magnitude of the complementary index $S_{\{-i\}}$. Although the computation of Correlation 2 is more expensive than Oracle 1 (based on $4N$ simulations), rLHDs could still be used to compute an averaged version of this estimator in the second stage of the adaptive strategy.

\medskip
Further work could be devoted to constructing optimized rLHDS to better fill the space of input variables. Going beyond rLHDs, a recent work has shown that two replicated orthogonal arrays (rOAs) are enough to estimate the whole set of first and closed second-order Sobol' indices by Oracle 2 \citep{Gilquin18}. One can thus reasonably expect that Oracle 1-like estimators could be implemented to limit the amount of spurious correlation in estimating small and moderate closed second-order indices. The adaptive strategy proposed in this paper could thus be extended to rOAs.

\section{Acknowledgments}

This work has been partly funded by the tripartite project devoted to Uncertainty Quantification, consisting of French Alternative Energies and Atomic Energy Commission (CEA), Electricity of France (EDF) and Framatome (FRA). The authors thank the two anonymous reviewers who contributed to improve the content of the paper.

\newpage

\appendix

\section{Comparison of estimators}
\label{appendix}


Throughout the appendix, suppose that both the mean $\mu_Y$ and variance $\sigma_Y^2$ of $Y$ are known. The starting point is to use the following decomposition of $y(.)$$:$
\begin{equation}
\label{expansion}
y(\mb{x})=\mu_{Y}+y_1(x_{i})+y_2(\mb{x}_{-i})+y_{12}(x_{i},\mb{x}_{-i}),
\end{equation}
where
\begin{itemize}
\item $y_1(x_{i})$ is the contribution of $x_i$ to the output,
\item $y_2(\mb{x}_{-i})$ is the contribution of all the other input variables $\mb{x}_{-i}$ to the output,
\item $y_{12}(x_{i},\mb{x}_{-i})$ is the contribution of the interaction between $x_{i}$ and $\mb{x}_{-i}$ to the output.
\end{itemize}
Based on Eq. \ref{expansion}, the Oracle 2 estimator is equal to
\begin{equation}
\label{oracle2Appendix}
\hat{S}^{Or2}_{\{i\}}= (N\sigma_Y^2)^{-1}\sum_{k=1}^{N} \big[y_1(x^k_{i})+y_2(\mb{x}^k_{-i})+y_{12}(x^k_{i},\mb{x}^k_{-i})\big]
\big[
y_1(x^k_{i})+y_2(\mb{w}^k_{-i})+y_{12}(x^k_{i},\mb{w}^k_{-i})\big]
\end{equation}
with $(N\sigma_Y^2)^{-1}\sum_{k=1}^{N} y_1(x^k_{i})^{2}$ converging to $S_{\{i\}}$ as $N\to\infty$, whereas every other term is \textit{spurious} correlation converging to $0$ as $N\to\infty$. However, the latter terms can badly affect the accuracy of (\ref{oracle2Appendix}) when $N$ is small or moderate. In \citet{Glen2012}, the amount of spurious correlation in Oracle 2 is presented as proportional to
\begin{equation} 
\label{spur2}
S_{\{-i\}}+\underline{S}_{\{i,-i\}}.
\end{equation} 
Eq. (\ref{spur2}) should be related to the total variance of the spurious terms present in Eq. (\ref{oracle2Appendix}). Let us focus on the two spurious terms 
\begin{equation}
\label{first_term}
(\sigma_Y^2)^{-1}y_2(\mb{x}_{-i})y_2(\mb{w}_{-i})
\end{equation}
and
\begin{equation}
\label{second_term}
(\sigma_Y^2)^{-1}y_{12}(x_{i},\mb{x}_{-i})y_{12}(x_{i},\mb{w}_{-i}).
\end{equation}
The variance of the sum of both Eqs. (\ref{first_term}) and (\ref{second_term}) is equal to the sum of the three contributions hereafter$:$
\begin{equation}
\vv\big[(\ref{first_term})]=(S_{\{-i\}})^{2},
\end{equation}
\begin{multline}
\vv\big[(\ref{second_term})]=
(\underline{S}_{\{i,-i\}})^{2}+
(\sigma_Y^4)^{-1}\text{Cov}(y_{12}(x_{i},\mb{x}_{-i})^{2},y_{12}(x_{i},\mb{w}_{-i})^{2})-\\
(\sigma_Y^4)^{-1}\ee\big[y_{12}(x_{i},\mb{x}_{-i})y_{12}(x_{i},\mb{w}_{-i})\big]^2
\end{multline}
and
\begin{equation}
2\,\text{Cov}\big[(\ref{first_term}),(\ref{second_term})\big]=
2(\sigma_Y^4)^{-1}\ee[y_2(\mb{x}_{-i})y_2(\mb{w}_{-i})y_{12}(x_{i},\mb{x}_{-i})y_{12}(x_{i},\mb{w}_{-i})].
\end{equation}
Thus, part of the variance is not equal to Eq. (\ref{spur2}), but instead to
\begin{equation}
\label{spur2_2}
(S_{\{-i\}})^2+(\underline{S}_{\{i,-i\}})^2.
\end{equation}
Moreover, Eq. (\ref{oracle2Appendix}) also includes other spurious terms omitted in \citet{Glen2012}. We can identify 
\begin{itemize}
\item two spurious terms comprising $y_1(.)$ and $y_2(.)$,
\item two spurious terms comprising $y_1(.)$ and $y_{12}(.)$,
\item two spurious terms comprising $y_2(.)$ and $y_{12}(.)$.
\end{itemize}
The variances and covariances of these terms depend on the model $y(.)$ and must also be taken into account in the calculation of the total variance of the spurious correlation affecting $\hat{S}^{Or2}_{\{i\}}$, which is thus not proportional to Eq. (\ref{spur2_2}). 

\medskip
In the same way, the Oracle 1 estimator can be expanded as$:$
\begin{multline}
\label{oracle1Appendix}
\hat{S}^{Or1}_{\{i\}}= (N\sigma_Y^2)^{-1}\sum_{k=1}^{N} \big[y_1(x^k_{i})+y_2(\mb{x}^k_{-i})+y_{12}(x^k_{i},\mb{x}^k_{-i})\big]
\big[
y_1(x^k_{i})+y_2(\mb{w}^k_{-i})+y_{12}(x^k_{i},\mb{w}^k_{-i})-\\
\big(y_1(w^k_{i})+y_2(\mb{w}^k_{-i})+y_{12}(w^k_{i},\mb{w}^k_{-i})\big)\big]
\end{multline}
We can see that the spurious correlation term depending on $(S_{\{-i\}})^{2}$ in Eq. (\ref{oracle2Appendix}) has vanished in Eq. \ref{oracle1Appendix}. However, the contribution of $(\underline{S}_{\{i,-i\}})^{2}$ is now doubled because of the two terms below$:$
\begin{equation}
\label{first_term_oracle1}
(\sigma_Y^2)^{-1} y_{12}(x_{i},\mb{x}_{-i})y_{12}(x_{i},\mb{w}_{-i})
\end{equation}
and
\begin{equation}
\label{second_term_oracle1}
-(\sigma_Y^2)^{-1} y_{12}(x_{i},\mb{x}_{-i})y_{12}(w_{i},\mb{w}_{-i}).
\end{equation}
A spurious correlation term proportional to $S_{\{i\}}$ appears through
\begin{equation}
\label{third_term_oracle1}
(\sigma_Y^2)^{-1} y_1(x_{i})y_1(w_{i}).
\end{equation}
According to \citet{Glen2012}, the spurious correlation tainting Oracle 1 is then proportional to 
\begin{equation}
\label{spur1}
S_{\{i\}}+2\underline{S}_{\{i,-i\}}
\end{equation}
or rather in reality to
\begin{equation}
\label{spur1_2}
(S_{\{i\}})^{2}+2(\underline{S}_{\{i,-i\}})^{2}.
\end{equation}
Once again, Eq. (\ref{spur1_2}) is only part of the variance of the sum of both Eqs. (\ref{first_term_oracle1}) and (\ref{second_term_oracle1}). Moreover, Eq. (\ref{oracle1Appendix}) includes
\begin{itemize}
\item two spurious terms comprising $y_1(.)$ and $y_2(.)$,
\item four spurious terms comprising $y_1(.)$ and $y_{12}(.)$,
\item two spurious terms comprising $y_2(.)$ and $y_{12}(.)$.
\end{itemize}
The rule given by \citet{Glen2012} that Oracle 1 outperforms Oracle 2 if
\begin{equation}
\label{conditionGlen}
S_{\{i\}}^{T}<\frac{1}{2}
\end{equation}
relies on the comparison between Eqs. (\ref{spur2}) and (\ref{spur1}). The previous arguments show this rule is in general theoretically wrong. However, suppose the special case where $S_{\{i\}}=S^{T}_{\{i\}}$. Then, Eq. (\ref{expansion}) is simplified to
\begin{equation}
y(\mb{x})=\mu_{Y}+y_1(x_{i})+y_2(\mb{x}_{-i})
\end{equation}
implying that the spurious correlation terms including $y_{12}(.)$ have vanished. It becomes possible to compare exactly the variance of Oracle 2 with that of Oracle 1. For Oracle 2, we have$:$
\begin{multline}
\label{varOr2add_1}
\ee\Big[\Big(y_1(x_{i})+y_2(\mb{x}_{-i})\Big)^2\Big(y_1(x_{i})+y_2(\mb{w}_{-i})\Big)^2\Big]=
\ee\Big[\Big(y_1(x_{i})^2+2y_1(x_{i})y_2(\mb{x}_{-i})+
y_2(\mb{x}_{-i})^2\Big)\\\Big(y_1(x_{i})^2+2y_1(x_{i})y_2(\mb{w}_{-i})+y_2(\mb{w}_{-i})^2\Big)\Big]
\end{multline}
Eq. (\ref{varOr2add_1}) is expanded as
\begin{multline}
\ee[y_1(x_{i})^4]+\ee[y_1(x_{i})^2]\ee[y_2(\mb{w}_{-i})^2]+\ee[y_1(x_{i})^2]\ee[y_2(\mb{x}_{-i})^2]+\ee[y_1(\mb{x}_{-i})^2]\ee[y_1(\mb{w}_{-i})^2]
\end{multline}
Thus,
\begin{equation}
\label{varO2}
\ee\Big[\Big(y_1(x_{i})+y_2(\mb{x}_{-i})\Big)^2\Big(y_1(x_{i})+y_2(\mb{w}_{-i})\Big)^2\Big]=\ee[y_1(x_{i})^4]+2\sigma_Y^4S_{\{i\}}S_{\{-i\}}+\sigma_Y^4 (S_{\{-i\}})^2
\end{equation}
Similarly for Oracle 1, we have
\begin{multline}
\label{varOr1add_1}
\ee\Big[\Big(y_1(x_{i})+y_2(\mb{x}_{-i})\Big)^2\Big(y_1(x_{i})-y_2(w_{i})\Big)^2\Big]=\\
\ee[y_1(x_{i})^4]+\ee[y_1(x_{i})^2]\ee[y_2(\mb{x}_{-i})^2]+\ee[y_1(w_{i})^2]\ee[y_2(\mb{x}_{-i})^2]+\ee[y_1(x_{i})^2]\ee[y_1(w_{i})^2]
\end{multline}
Thus,
\begin{equation}
\label{varO1}
\ee\Big[\Big(y_1(x_{i})+y_2(\mb{x}_{-i})\Big)^2\Big(y_1(x_{i})-y_2(w_{i})\Big)^2\Big]
=\ee[y_1(x_{i})^4]+2\sigma_Y^4S_{\{i\}}S_{\{-i\}}+\sigma_Y^4 (S_{\{i\}})^2
\end{equation}
By comparing Eqs. (\ref{varO2}) with (\ref{varO1}), we can conclude that Oracle 1 is more accurate than Oracle 2 if and only if
\begin{equation}
\label{conditionGlenAdditive}
S_{\{i\}}<\frac{1}{2}.
\end{equation}

\bibliographystyle{elsarticle-harv}

\bibliography{biblio}

\end{document}